%% file: paper.short.II.tex
\documentclass[usenatbib]{mn2e}
\usepackage{amsmath}
\usepackage{natbib}
\usepackage{epsfig}
\usepackage{txfonts}

\input{Befehle}

\newcommand{\expf}[1]{{{\rm e}^{#1}}}
\newcommand{\Teb}{\bar{T}_{\rm e}}

\newcommand{\MJy}{{\rm MJy}}
\newcommand{\sr}{{\rm sr}}
\newcommand{\keV}{{\rm keV}}

\usepackage{color}

\usepackage{hyperref}
\usepackage{grffile}
\usepackage{graphics}
\usepackage{booktabs}

\title[Multiple scattering SZ II]
{Multiple scattering Sunyaev-Zeldovich signal II: relativistic effects}

\author[Chluba \& Dai]{
J.~Chluba\thanks{E-mail: jchluba@pha.jhu.edu} and 
L. Dai\thanks{E-mail:ldai@pha.jhu.edu}
\\
Department of Physics and Astronomy, 
Johns Hopkins University, Bloomberg Center, 
3400 N. Charles St., Baltimore, MD 21218, USA
}

\begin{document}

\date{{Accepted November 22. Received 2013 September 12}}

\maketitle

\begin{abstract}
We study the multiple scattering Sunyaev-Zeldovich (SZ) signature, extending our previous analysis to high-temperature clusters. 
We consistently treat the anisotropy of the ambient radiation field caused by the first scattering and also consider lowest order kinematic terms.
We show that due to temperature corrections monopole through octupole anisotropy of the singly scattered SZ signal attain different spectra in the second scattering. The difference becomes more pronounced at high temperature, and thus could be used to constrain {\it individual} line of sight {\it moments} of the electron density and temperature profiles.
While very challenging from the observational point of view, this further extends the list of possible SZ observables that will be important for 3D cluster-profile reconstruction, possibly helping to break geometric degeneracies caused by projection effects.
We also briefly discuss the scattering of primordial CMB anisotropies by SZ clusters.
\end{abstract}


\section{Introduction}
\label{sec:Intro}
For the interpretation of future high-resolution and high-sensitivity SZ measurement for individual clusters, obtained with, e.g.,  ALMA\footnote{Atacama Large Millimeter/submillimeter Array},
CARMA\footnote{Combined Array for Research in Millimeter-wave
Astronomy}, CCAT\footnote{Cornell Caltech Atacama Telescope},
and MUSTANG\footnote{MUltiplexed Squid TES Array at 90 GHz},
it is important to understand the precise dependence of the SZ signal on cluster parameters.
In this work, we extend our recent analysis \citep[][CDK13 hereafter]{Chluba2013mSZ} of the multiple scattering SZ signal to high-temperature cluster atmospheres. For references and more general introduction to the problem we refer to CDK13.
The main improvement is that previous works \citep{Dolgov2001, Itoh2001, Colafrancesco2003, Shimon2004} neglected the scattering-induced anisotropy of the singly scattered radiation field, but as shown in CDK13, this approximation is rather crude, even at the lowest order in the electron temperature $\Te$. 

At higher temperature, additional significant corrections arise. Most importantly, in the second scattering the spectra of the singly scattered monopole through octupole radiation anisotropies are affected in different ways by energy exchange with the moving electrons. This effect might allow separating each contribution using future high-resolution, high-sensitivity SZ data of individual clusters, and thus extends the list of possible SZ observables. 
We furthermore compute the lowest order kinematic correction caused by the second scattering. The signal depends on {\it all three} components of the cluster's peculiar motion, and thus could in principle be used to constrain the cosmological, large-scale velocity field.

As another example for the scattering of anisotropic cosmic microwave background (CMB) radiation by hot clusters, we calculate the contributions of primordial CMB temperature fluctuations.
In addition to the thermal SZ \citep[thSZ;][]{Zeldovich1969} and kinetic SZ \citep[kSZ;][]{Sunyaev1980} signal, clusters act as an {\it optical depth screen}, imprinting additional small-scale temperature fluctuations (with a thermal spectrum) on to the large-scale CMB anisotropies \citep[see][for related discussion]{Zeldovich1980b, Sunyaev1981, Carlos2010}; however, as we show here, this effect remains unobservable at this stage. 
Also, the hot cluster electrons exchange energy with photons drawn from the primordial CMB dipole, quadrupole and octupole anisotropy. Again, this effect is small, but it could be noticeable close to the SZ null. If a large primordial dipole were present, future samples of clusters can thus in principle be used to place a limit on its amplitude with stacking techniques.

\section{Second scattering at high temperatures}
\label{sec:multiple_scatt}
In this section we generalize the calculation of CDK13, providing a quasi-exact treatment of the second scattering including higher order temperature corrections. We start our analysis with the isothermal case, and then derive the expressions for general geometry of the intra cluster medium (ICM).
Most technical details are presented in the appendices. In particular, there we discuss the {\it scattering kernel} describing the interaction of low-energy CMB photons with hot electrons for different multipoles of the radiation field. While not the central part of the paper, these expressions might be of more general interest. We follow the same notation as CDK13.

The analysis of CDK13 considered the multiple scattering SZ effect at lowest order in the temperature parameter $\The=k\Te/\me c^2$. Here, we evaluate the full Compton collision term, $\mathcal{C}[n]$, adding temperature corrections until the result converges. This changes both the singly scattered as well as the doubly scattered SZ signals.
Explicit expressions for the collision term, valid for $\Te\gg \Tg$, were given in \citet[][CNSN hereafter]{ChlubaSZpack}. We provide a brief summary of the most relevant equations in Appendix~\ref{app:collision_term}.

\subsection{SZ signal in the single scattering limit}
As shown in \citet[][CSNN hereafter]{Chluba2012moments}, the thSZ signal introduced by the first scattering event is given by
\beal
\label{eq:dist_single}
\Delta  I^{(1)} (x, \vgh)  \approx  
 S^{(0)}(x, \tau, \Teb)+\sum_{k=1}^\infty S^{(k+1)}(x, \tau, \Teb) \, \omega^{(k)}.
\end{align}
Here, we introduced the frequency variable $x=h\nu/kT_0$ (with the CMB monopole temperature $T_0=2.726\,$K and the Planck and Boltzmann constants $h$ and $k$, respectively). The SZ signal function, $S^{(0)}(x, \tau, \Teb)$, describes the effect of Compton scattering on the CMB monopole spectrum for electrons at SZ-weighted temperature $\Teb=\tau^{-1}\int \Te \id \tau'$ with total line of sight optical depth $\tau(\vgh) = \int\Ne(\vek{r}) \sigT \id l$, where $\Ne(\vek{r})$ denotes the free electron profile. At the lowest order of the electron temperature, $S^{(0)}(x, \tau, \Teb)$ simply describes the standard thSZ effect.
Using the Compton kernel, $\mathcal{P}_0(s, \Te)$, to describe the scattering of the radiation monopole from frequency $\nu'$ to $\nu$ with $s=\ln(\nu'/\nu)$ [see Appendix~\ref{sec:collision_term_kernel} for more details], we may also write 
\beal
\label{eq:S0_def}
S^{(0)}(x, \tau, \Teb)= x^3 I_{\rm o} \tau \int \mathcal{P}_0(s, \Teb) [ \nbb(x\expf{s})-\nbb(x) ] \id s,
\end{align}
where the constant $I_{\rm o}=(2h/c^2)(kT_0/h)^3\approx 270\,\MJy\,\sr^{-1}$ and $\nbb(x)=(\expf{x}-1)^{-1}$.
The second term in Eq.~\eqref{eq:dist_single} accounts for the variation of the electron temperature along the line of sight, with temperature moments $\omega^{(k)}=\tau^{-1} \int (\Te/\Teb-1)^{k+1} \id \tau'$. It really follows from a simple Taylor series of $S^{(0)}(x, \tau, \Te)$ around $\Teb$ before the line of sight average is carried out.
The dominant contribution is from\footnote{The first derivative term vanishes when expanding around $T=\Teb$.} $k=1$, related to the line of sight {\it dispersion} of the electron temperature, while higher order moments typically drop rapidly (cf. CSNN). 
The functions $S^{(k)}(x, \tau, \Te)\equiv (\Te^k/k!)\,\partial^k_{\Te} S^{(0)}(x, \tau, \Te)$ can be computed with high precision using {\sc SZpack} (see CNSN) over a wide range of temperatures ($\Te\lesssim 75\,\keV$) and frequencies ($0.01\lesssim x\lesssim 30$). In particular, {\sc SZpack} overcomes the convergence issues of previous approaches without increasing the computational burden. 
In terms of an asymptotic expansion for small $\The\equiv k\Te/(\me c^2)$, we can also express $S^{(0)}(x, \tau, \Te)$ as
\beal
\label{eq:dist_S0_approx}
S^{(0)}(x, \tau, \Te)\approx x^3 I_{\rm o} \tau \,\The \sum_{k=0}^{k_{\rm max}} Y_k(x) \, \The^{k}.
\end{align}
The functions $Y_k(x)$ can be found in \citet{Itoh98} or CNSN up to $k_{\rm max}=10$. Setting $k_{\rm max}=0$ gives the standard thSZ formula, with $Y_0(x)=\frac{x\expf{x}}{(\expf{x}-1)^2}[x\coth(x/2)-4]$ \citep{Zeldovich1969}.

We note that $S^{(k)}(x, \tau, \Te)\equiv \tau \, \hat{S}^{(k)}(x, \Te)$. This factorization can be directly deduced from Eq.~\eqref{eq:dist_S0_approx}, and is a consequence of the single scattering limit. This property is important for deriving the second scattering correction for isothermal ICM.

\subsection{Second scattering correction: isothermal case}
\label{sec:second_scatt}
To obtain the second scattering correction, we simply have to reapply the Compton collision term to the singly scattering radiation field seen at every location $\vek{r}$ inside the cluster. The observed second scattering signal in the direct $\vgh$ is then given by averaging all contributions along the line of sight.
This approach is possible, because the typical optical depth of clusters is very small allowing a Born-series expansion of the problem (see CDK13).
For an isothermal distribution of free electrons ($\omega^{(k)}=0$) with general electron density profile $\Ne(\vek{r})$, around the location $\vek{r}$ and in the direction $\vghp$, from Eq.~\eqref{eq:dist_single} we have $\Delta  I^{(1)} (x, \vek{r}, \vghp)\approx S^{(0)}(x, \tau(\vek{r}, \vghp), \Te)\equiv  \tau(\vek{r}, \vghp)\, \hat{S}^{(0)}(x, \Te)$. 
This expression shows that the anisotropy of the local radiation field is solely determined by variations of the optical depth, $\tau(\vek{r}, \vghp)$ around $\vek{r}$. Using the scattering kernel, $\mathcal{P}_\ell(s, \Te)$, for different radiation multipoles $\ell$, the full Compton collision term $\mathcal{C}[n]$ is given by Eq.~\eqref{eq:Boltzmann_kernel}. 
Introducing the distortion of the photon occupation number $\hat{s}^{(0)}(x, \Te)=\hat{S}^{(0)}(x, \Te)/(x^3 I_{\rm o})$ caused by the first scattering event and inserting into $\mathcal{C}[n]$ then yields the second scattering SZ signal observed in the direction $\vgh$
\beal
\label{eq:dist_second}
\Delta  I^{(2)} (x, \vgh) & =\Delta  I^{(2), \rm T} (x, \vgh)+\Delta  I^{(2), \sigma} (x, \vgh)+\Delta  I^{(2), \rm E} (x, \vgh),
\\[1mm] \nonumber
\Delta  I^{(2), \rm T}  (x, \vgh)
&= \hat{S}^{(0)}(x, \Te) \left[ \left<\tau_0\right> + \frac{\left<\tau_2\right>}{10} - \frac{\tau^2}{2} \right],
\nonumber\\
\nonumber
\Delta  I^{(2), \sigma} (x, \vgh)
&=
\hat{S}^{(0)}(x, \Te)
\sum_{\ell=1}  \left<\tau_{\ell}\right> \left[\int \mathcal{P}_\ell(s, \Te) \id s -\frac{\delta_{\ell 2}}{10} \right],
\\ \nonumber
\Delta  I^{(2), \rm E} (x, \vgh)
&=x^3 I_{\rm o} \sum_\ell  \left<\tau_{\ell}\right> \!
\int \!\mathcal{P}_\ell(s,  \Te)\left[\hat{s}^{(0)}(x\,\expf{s}, \Te)- \hat{s}^{(0)}(x, \Te)\right] \!\id s,
\\ \nonumber
\tau_\ell(\vek{r})&=\frac{2\ell+1}{4\pi}\int  \id^2\vghp P_\ell(\vgh\cdot\vghp) \,\tau(\vek{r}, \vghp),
\\ \nonumber
 \left<\tau_\ell\right>&=\int \tau_\ell(\vek{r}) \sigT \Ne(\vek{r}) \id l =\int \tau_\ell(\vek{r}) \id \tau \propto \tau^2/2.
\end{align}
%
Here, 
$P_\ell(x)$ is the Legendre polynomial and $\left<\tau_{\ell}\right>$ denotes the weighted (by $\Ne(\vek{r})\sigT$) average optical depth Legendre coefficient integrated along the line of sight. 
We separated terms that leave the singly scattered spectrum unchanged (first two terms) from those that alter the spectrum (last term). 
At this point, we only assumed that the scattering medium is isothermal with temperature much larger than the CMB temperature, so that Eq.~\eqref{eq:dist_second} gives a quasi-exact representation of the second scattering signal.

The first term of Eq.~\eqref{eq:dist_second}, $\Delta  I^{(2), \rm T}$, is simply the Thomson scattering correction to the thSZ signal.
Comparing with Eq.~(9b) of CDK13, shows that only $\Delta  I^{(1)}/\tau\approx \The x^3 I_{\rm o} Y_0(x)$ was replaced with $\hat{S}^{(0)}(x, \Te)$ to account for all higher temperature corrections to the singly scattered radiation field.
The second term of Eq.~\eqref{eq:dist_second}, $\Delta  I^{(2), \sigma}$, is caused by temperature-dependent corrections to the total scattering cross section of each radiation multipole. At the lowest order of $\Te$, CDK13 showed that this term is negligible, thus even for higher temperature one expects $\Delta  I^{(2), \sigma}$ only gives a very small correction to $\Delta  I^{(2), \rm T}$. Defining $\Delta \sigma_\ell=\int \mathcal{P}_\ell(s, \Te) \id s -\delta_{\ell 2}/10$, we can compute the temperature-dependent terms of the total scattering cross section numerically. 
%
\begin{figure}
\centering
\includegraphics[width=\columnwidth]{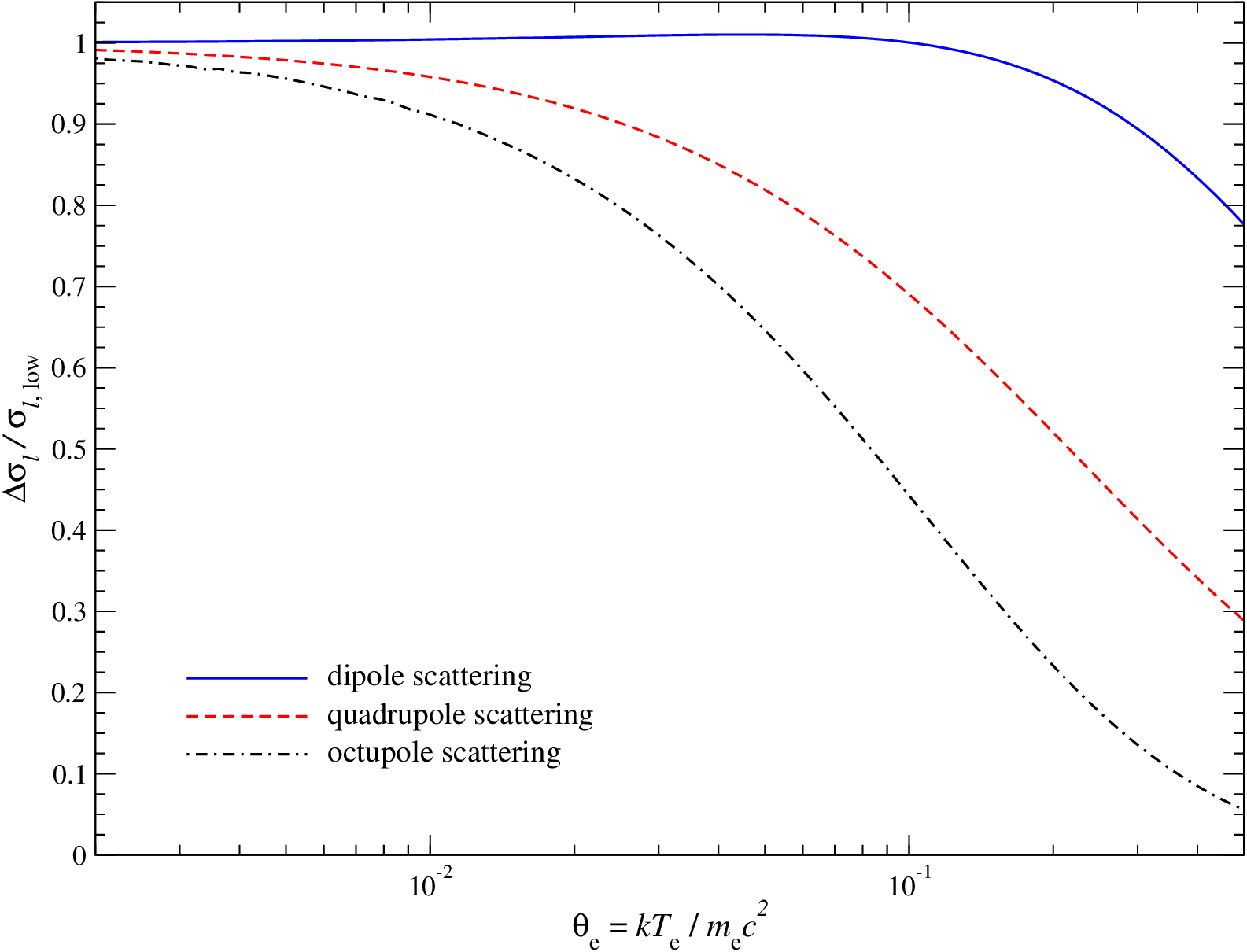}
\caption{Temperature correction to the scattering cross section for dipole through octupole anisotropy. We normalized by the lowest order correction, $\sigma_{1, \rm low}=-(2/5)\sigT \The$, $\sigma_{2, \rm low}=-(3/5)\sigT \The$ and $\sigma_{3, \rm low}=(6/35)\sigT \The$. 
}
\label{fig:sig_temp}
\end{figure}
The results are shown in Fig.~\ref{fig:sig_temp}. The cross section correction for scattering of the dipole anisotropy is well described by the leading order term, $\Delta \sigma_1\simeq -(2/5)\sigT \The$, up to $k\Te\sim 50\,\keV$. For the quadrupole and octupole higher order temperature corrections need to be included to give an accurate representation of the cross section correction, but in general the contribution to the SZ signal remains negligibly small so that below we omit it.

Equation~\eqref{eq:dist_second} also shows that for isothermal electrons, the spectral (redistribution) part can be computed without knowledge of the optical depth anisotropies. This means that even for general geometries or electron density profiles, the integrals over scattering angles and electron momenta can be carried out independently. 
Another simplification is possible, because multipoles with $\ell>3$ scatter at leading order $\mathcal{O}(\The^{\ell-2})$ in temperature (see CDK13). One can furthermore expect that the high $\ell$ moments of the $y$-parameter are smaller than those for $\ell\leq 3$. Hence, only contributions from multipoles $\ell\leq 3$ need to be considered, even for very hot clusters, $k\Te \simeq 25\,\keV$. 
These aspects greatly simplify the calculation, especially when using simulated clusters to obtain more realistic predictions for the second scattering SZ signal.

\subsubsection{Second scattering correction to the thSZ effect in the Fokker-Planck approximation}
\label{sec:Fokker_planck_second_scattering}
To compute the second scattering correction caused by energy exchange with thermal electrons, $\Delta  I^{(2), \rm E}(x, \vgh)$, we have to carry out the integral over the scattering kernels, giving a quasi-exact result for the correction. Numerically, this is straightforward but for analytic estimates and at low temperatures it is useful to give the corresponding terms using a Fokker-Planck approximation of the collision term. With Eq.~\eqref{eq:dist_second_rewrite} we find
\beal
\label{eq:dist_second_rewrite_bb}
\Delta  I^{(2), \rm E} (x, \vgh)
&\approx  x^3 I_{\rm o} \sum_{\ell=0}  \left<\tau_{\ell}\right> \sum_{k=1} I^{k}_\ell(\Te) \, x^{k}\partial^k_x \,\hat{s}^{(0)}(x, \Te).
\end{align}
For the dipole and quadrupole all kernel moments $I^{k}_\ell(\Te)$ up to $\mathcal{O}(\The^9)$ were provided by CNSN. For the octupole they are summarized in Table~\ref{tab:oct}. 
Equation~\eqref{eq:dist_second_rewrite_bb} can be further simplified (see Appendix~\ref{app:FP_result}) into the compact form 
\beal
\label{eq:result_expansion_DI2E_l}
\Delta  I^{(2), \rm E} (x, \vgh)& \approx \sum_{\ell=0} \left<\tau_{\ell}\right> \Delta I^{(2), \rm E}_\ell,
\\[1mm]
\Delta I^{(2), \rm E}_\ell(x)
&=x^3 I_{\rm o} \int \!\mathcal{P}_\ell(s,  \Te)\left[\hat{s}^{(0)}(x\,\expf{s}, \Te)- \hat{s}^{(0)}(x, \Te)\right] \!\id s
\nonumber\\[-0.5mm] \nonumber
&\approx \The^2 x^3 I_{\rm o} \sum^{n-1}_{k=0}\The^k \, Y^{(\ell)}_k(x),
\end{align}
where the distortion functions $Y^{(\ell)}_k(x)$ are defined by Eq.~\eqref{eq:result_expansion_DI2E_l_app}.
CDK13 already showed that at the lowest order of the electron temperature, $Y^{(\ell)}_0(x)\propto Y^{(0)}_0(x)\equiv Y^\ast_0(x)$ [also defined in CDK13]; however, at higher order in temperature the independent multipoles scatter differently:
\beal
Y^{(0)}_1
\!&=\!
\left[80 D_x +590 D^2_x +\frac{3492}{5} D^3_x+\frac{1271}{5} D^4_x+\frac{168}{5} D^5_x+ \frac{7}{5} D^6_x\right]\nbb,
\nonumber\\[1mm]
Y^{(1)}_1
\!&=\!
 -\frac{2}{5}\left[88 D_x +697 D^2_x +\frac{4206}{5} D^3_x+\frac{3081}{10} D^4_x+\frac{204}{5} D^5_x+ \frac{17}{10} D^6_x\right]\nbb,
\nonumber\\[1mm]
Y^{(2)}_1
\!&=\!
 \frac{8}{7}\left[D_x +58 D^2_x +\frac{2142}{25} D^3_x+\frac{6643}{200} D^4_x+\frac{447}{100} D^5_x+ \frac{149}{800} D^6_x\right]\nbb,
\nonumber\\[1mm]
Y^{(3)}_1
\!&=\!
\frac{24}{5}\left[D_x -\frac{17}{7} D^2_x -\frac{216}{35} D^3_x-\frac{739}{280} D^4_x-\frac{51}{140} D^5_x-\frac{17}{1120} D^6_x\right]\nbb,
\nonumber
\end{align}
with $D^k_x=x^k \partial^k_x$. The derivatives of the Planckian spectrum for any $k$ can be calculated in closed form using Eulerian numbers, as explained in Appendix~A of CNSN. Higher order temperature terms can be obtained in a similar way, but since the formulae are not very illuminating we {\it omit} them here. 
Up to $\mathcal{O}(\The^{11})$ all correction terms for $\ell\leq 3$ are available for {\sc SZpack}; however, the convergence of these expressions is very limited, as discussed below.
The functions $Y^{(0)}_0$ and $Y^{(1)}_0$ correspond to $Z_0$ and $Z_1$ of \citet{Itoh2001}, while the other $Y^{(\ell)}_k$ did not appear in the literature before but are required to describe the second scattering signal correctly.

\begin{figure}
\centering
\includegraphics[width=\columnwidth]{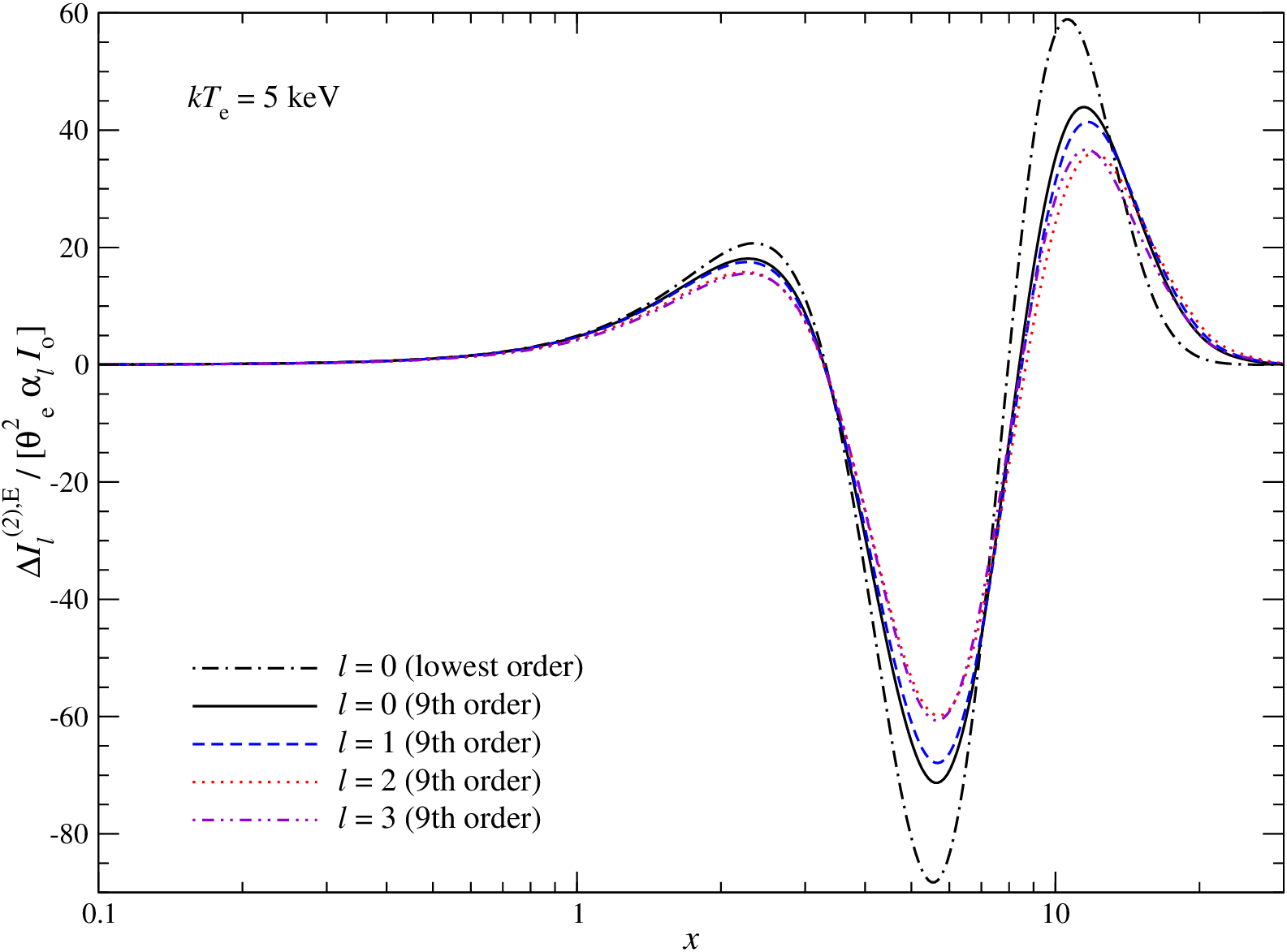}
\\[2mm]
\includegraphics[width=\columnwidth]{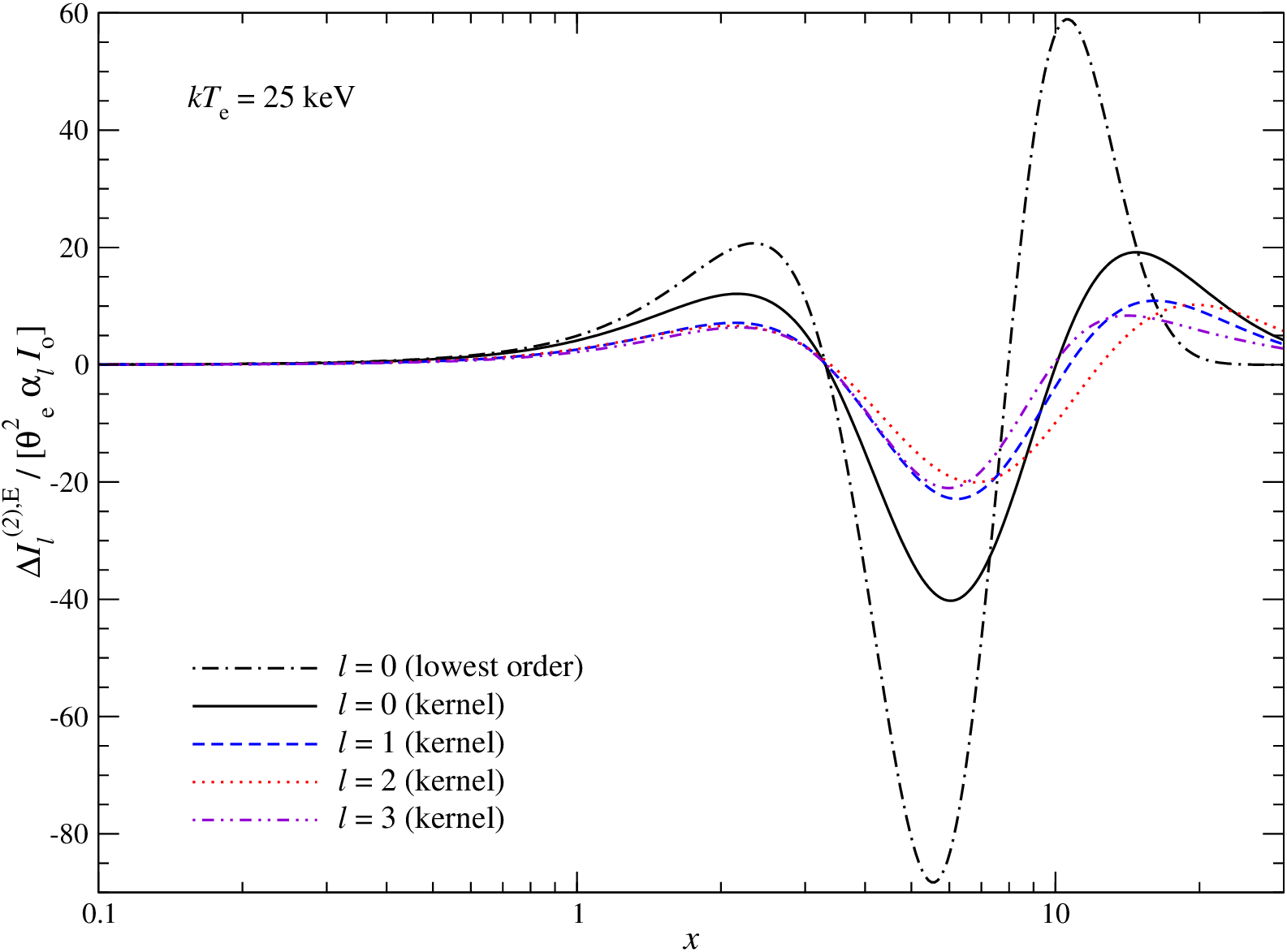}
\caption{Second scattering correction, $\Delta  I^{(2), \rm E}_\ell$, for different multipoles in comparison with the lowest order expansion, $\Delta  I^{(2), \rm E}_\ell\approx \alpha_\ell \, x^3 I_{\rm o} \The^2 Y^{(0)}_0(x)$ [cf. Eq.~\eqref{eq:result_expansion_DI2E_l}], with $\alpha_0=1$, $\alpha_0=-2/5$, $\alpha_2=1/10$, and $\alpha_0=-3/70$. 
We rescaled all curves by $\The^2 \alpha_\ell$ to make them comparable. For $k\Te=5\,\keV$ the results are computed using the asymptotic expansion up to $\mathcal{O}(\The^{11})$ [$9^{\rm th}$ order correction in $\The$], while for $k\Te=25\,\keV$ the kernel approach was applied. At higher temperatures the differences between the scattering signals for independent multipoles become more pronounced. 
}
\label{DI2_E.Te_5keV_25keV}
\end{figure}
\subsubsection{Temperature-dependence for different multipoles}
CDK13 showed that at the lowest order in $\The$, the spectral distortion functions $\Delta  I^{(2), \rm E}_\ell$ for $\ell\leq 3$ all look the same modulo a constant coefficient. However, already at rather low temperatures $k\Te\simeq 5\,\keV$, the lowest order expansion $\propto Y^{(\ell)}_0$ becomes inaccurate, and higher order temperature corrections need to be included. This is shown in Fig.~\ref{DI2_E.Te_5keV_25keV} for $\ell\leq 3$. Not only $\Delta  I^{(2), \rm E}_0$ differs significantly from the lowest order expression, but also the dependence of $\Delta  I^{(2), \rm E}_\ell$ on multipole is noticeable.
At low temperatures, the asymptotic expansion up to $\mathcal{O}(\The^{11})$ converges very well, but already at $k\Te\simeq 8\,\keV-10\,\keV$ the series breaks down. For $k=0$ this was also found by \citet{Itoh2001} and \citet{Dolgov2001} using the isotropic scattering approximation (ISA). 

At higher temperatures numerical integration with the Compton kernel (see Appendix~\ref{sec:collision_term_kernel}) should be used to obtain accurate results. 
This is because a Fokker-Planck expansion becomes non-perturbative, converging asymptotically slowly\footnote{We find that the convergence rate is even slower than for the singly scattered thSZ effect, for which the asymptotic expansion works up to electron temperatures $k\Te \simeq 13\keV$ (see CSNN).}.
As Fig.~\ref{DI2_E.Te_5keV_25keV} illustrates, in this case the differences between the scattering signals for independent multipoles become more pronounced. The largest difference is found at high frequencies, in the Wien tail of the CMB. 
This effect could in principle allow separation of the different optical depth moments for $\ell\leq 3$ with future SZ measurements.

\begin{figure}
\centering
\includegraphics[width=\columnwidth]{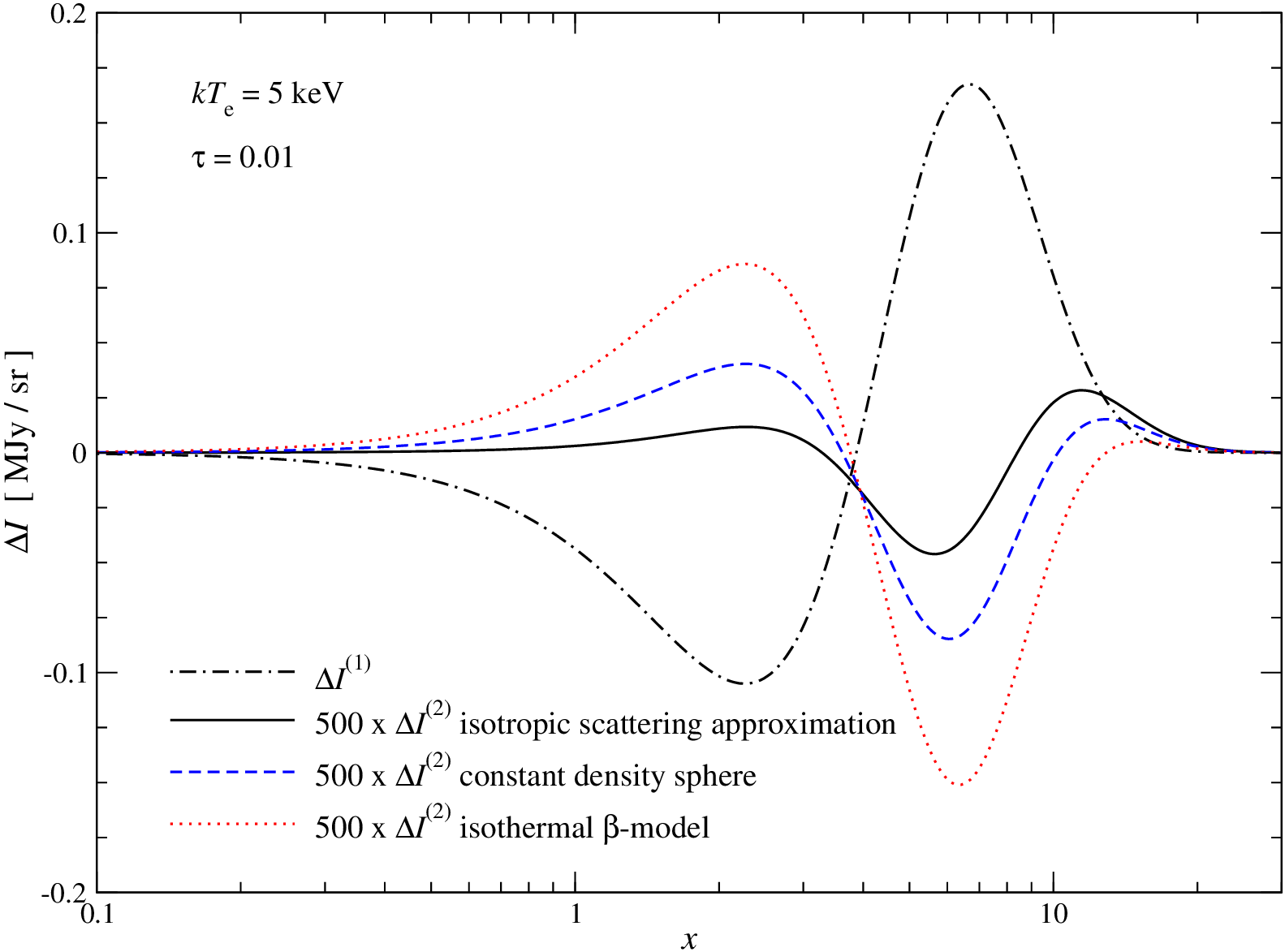}
\\[2mm]
\includegraphics[width=\columnwidth]{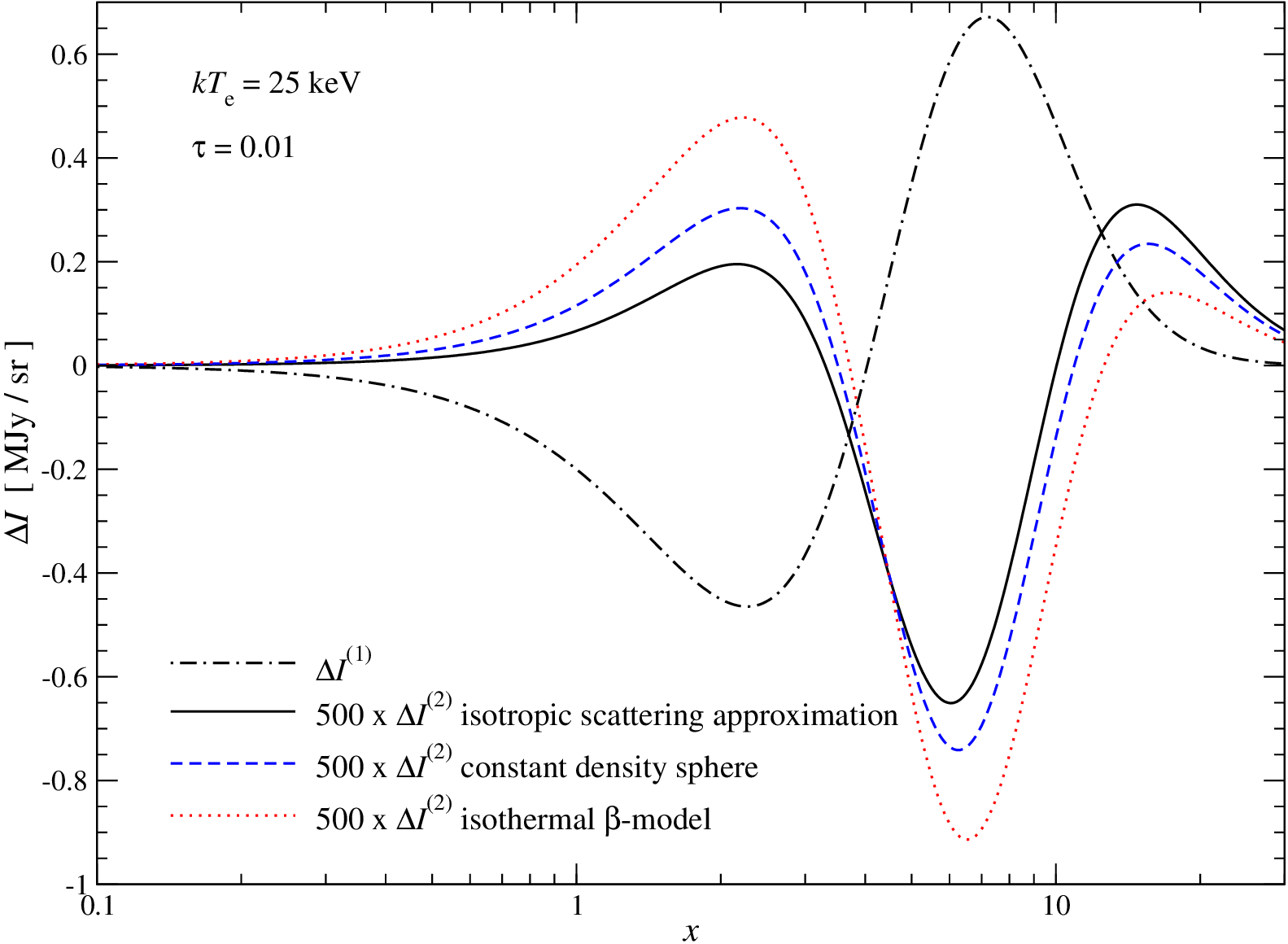}
\caption{Total second scattering correction for the constant density, isothermal sphere and the isothermal $\beta$-model at high temperatures [see CDK13 for details on the computation of $\left<\tau_\ell\right>$]. For comparison, we show the singly scattering thSZ effect which is the same in both cases. Also, in the ISA, the second scattering SZ signal is independent of the geometry; however, due to Thomson scattering corrections, the second scattering signal directly probes the geometry of the electron distribution. At higher temperatures this effect becomes less important.}
\label{fig:SZ_signal_multi-scatt.Te_5keV}
\end{figure}
\subsubsection{Total second scattering signal and comparison to the ISA}
To illustrate the amplitude of the second scattering contribution at higher temperatures,  in Fig.~\ref{fig:SZ_signal_multi-scatt.Te_5keV} we show the total signal for different temperatures. In our formulation, the ISA reads 
\beal
\label{eq:ISA_high_temp}
\Delta  I^{(2), \rm E, iso}
&= \frac{\tau^2}{2} \Delta I^{(2), \rm E}_0(x)
\approx x^3 I_{\rm o} \frac{\The^2 \tau^2}{2} \sum^{k_{\rm max}}_{k=0} \The^k Y^{(0)}_k.
\end{align}
The ISA does not account for the largest correction caused by Thomson scattering. Its relative contribution becomes smaller at higher temperatures (see the lower panel of Fig.~\ref{fig:SZ_signal_multi-scatt.Te_5keV}); however, it clearly affects the overall characteristic of the second scattering signal. We can also notice the effect of geometry on the SZ signal, an effect that also is not captured by the ISA. 

The total correction reaches $\simeq 0.2\%$ for $k\Te = 25\keV$ at intermediate frequencies, while in the Wien tail ($x\gtrsim 10$) it exceeds this level. Close to the crossover frequency the contribution from second scattering terms is also significant, shifting its position.
Although the total correction differs significantly from the one found in previous works, the second scattering SZ signal remains small. Even the shift in the position of the crossover frequency is small and strongly degenerate with other effects in this band, e.g., because of higher order temperature correction to the thSZ, line of sight temperature variations, or the kSZ effect.
Nevertheless, it in principle opens another way to constrain the distribution of electrons inside the cluster and in the future it might help with the 3D cluster-profile reconstruction.

\subsection{Second scattering correction: non-isothermal case}
\label{sec:multiple_scatt_non_iso}
CDK13 showed how spatial variations in the electron temperature affect the SZ signal at the lowest order in $\Te$.
To include this effect at higher order, we can proceed in a similar way. Because more detailed modelling of the ICM is required to account for spatial electron temperature variations, here we just outline the procedure, emphasizing the main dependences on the cluster's atmosphere, i.e. moments of $\Ne(\vek{r})$ and $\Te(\vek{r})$. A simulation-based analysis is left for the future.

Starting with Eq.~\eqref{eq:dist_single}, it is clear that in the general case the anisotropy of the singly scattered radiation field also depends on the variations of $\Teb(\vek{r}, \vgh)$ and $\omega^{(k)}(\vek{r}, \vgh)$. For the singly scattered radiation field, the SZ-weighted temperature $\Teb=\tau^{-1} \int \Te \id \tau'$ is important; however, in the second scattering the weighting changes by another factor of $\tau$ to $\Teb^\ast=2\tau^{-2} \int \tau' \,\Te \id \tau'$. This can be seen when considering the scattering of photons out of the line of sight. Neglecting moments of the temperature field with powers $k>1$ we can express the singly scattered radiation field at location $\vek{r}$ as
\beal
\label{eq:sing_non_iso}
\Delta  I^{(1)} (x, \vghp, \vek{r})  &\approx  
\tau(\vghp, \vek{r})\left[ 
\hat{S}^{(0)}(x, \Teb^\ast)\, 
+\hat{S}^{(1)}(x, \Teb^\ast) \, \Theta(\vghp, \vek{r})\right.
\nonumber\\
&\qquad\qquad\qquad \qquad\left.+\hat{S}^{(2)}(x, \Teb^\ast) \, \omega^{(1)}(\vghp, \vek{r})\right],
\end{align}
where $\Theta(\vghp, \vek{r})=\Teb(\vghp, \vek{r})/\Teb^\ast-1$ and $\omega^{(1)}=\tau^{-1}\int \Theta(\vghp, \vek{r})^2\id\tau$. Defining $\Teb^\ast=2\tau^{-2} \int \tau' \,\Te \id \tau'$ makes the second term vanish once the line of sight integral for the second scattering is carried out. This yields
\beal
\Delta  I^{(2)}_- (x, \vgh)  &\approx  
\frac{\tau^2(\vgh)}{2}\left[ 
\hat{S}^{(0)}(x, \Teb^\ast)\, 
+\hat{S}^{(2)}(x, \Teb^\ast) \, \omega^{(1),\ast}(\vgh)\right],
\nonumber\\
\omega^{(1),\ast}(\vgh)&=\frac{2}{\tau^2}\int \tau' \omega^{(1)}(\vek{r})\id \tau'.
\end{align}
Although the second scattering out of the line of sight does not include any energy exchange, the singly scattered spectrum, being a mixture of SZ signals from scattering electrons with different temperatures, is reweighed slightly, causing a small change in the shape of the average photon distribution. The change from $\Teb$ to $\Teb^\ast$ is most important, but also the effective second temperature moment, $\omega^{(1),\ast}$, in general differs from $\omega^{(1)}$. This small effect is absent at the lowest order in the electron temperature.

To derive the intensity change caused by scatterings back into the line of sight, we insert Eq.~\eqref{eq:sing_non_iso} into the collision integral. To give the final expression we need expansions for $\tau(\mu) \Theta(\mu)=\sum_{\ell=0} [\tau\Theta]_\ell P_\ell(\mu)$ and $\tau(\mu)  \omega^{(1)}(\mu)=\sum_{\ell=0} [\tau \omega^{(1)}]_\ell P_\ell(\mu)$, where we defined the Legendre coefficients $[X]_\ell\equiv \frac{2\ell+1}{2}\int X(\mu) P_\ell(\mu)\id \mu$. We furthermore introduce
\beal
\label{eq:DI_spectral_integral}
\Delta \hat{S}_{\ell}(\Te, \Teb^\ast)&=x^3 I_{\rm o} \!\!\int_{-\infty}^\infty\!\! \!\mathcal{P}_\ell(s, \Te) \left[ \hat{s}^{(0)}(x \expf{s}, \Teb^\ast)-\hat{s}^{(0)}(x, \Teb^\ast)\right] \! \!\id s,
\end{align}
where it is important that for the redistribution of photons at location $\vek{r}$ along the line of sight the local electron temperature $\Te(\vek{r})$ is relevant. With this we  find
\beal
\label{eq:non_iso_DI2}
&\Delta  I^{(2)}(x, \vgh)  =\Delta  I^{(2)}_+(x, \vgh) -\Delta  I^{(2)}_-(x, \vgh) 
\\\nonumber
&\;
\approx  
\left[ \left<\tau_{0}\right>+\frac{\left<\tau_{2}\right>}{10}
- \frac{\tau^2}{2} \right] \hat{S}^{(0)}(x, \Teb^\ast)
+\left[ \left<[\tau\Theta]_{0}\right>+\frac{\left<[\tau\Theta]_{2}\right>}{10} \right] \hat{S}^{(1)}(x, \Teb^\ast)
\nonumber\\
&\;\quad
+\left[ \left<[\tau\omega^{(1)}]_{0}\right>+\frac{\left<[\tau\omega^{(1)}]_{2}\right>}{10}
- \frac{\tau^2}{2}\omega^{(1),\ast} \right] \hat{S}^{(2)}(x, \Teb^\ast)
\nonumber\\
&\quad\quad+\left<\sum_{\ell=0}\left[\tau_{\ell}
+[\tau\Theta]_{\ell} \Teb^\ast\partial_{\Teb^\ast} 
+\frac{1}{2} [\tau\omega^{(1)}]_\ell (\Teb^\ast)^2\partial^2_{\Teb^\ast} \right]\,\Delta \hat{S}_{\ell}(\Te, \Teb^\ast)\right>,
\nonumber
\end{align}
where $\left<X\right>\equiv \int X(\vek{r}) \id \tau$ as before. The last line of sight average can be further simplified by inserting
\beal
\label{eq:DS_expansion}
\Delta \hat{S}_{\ell}(\Te, \Teb^\ast)&\approx \Delta \hat{S}_{\ell}(\Te^\ast, \Teb^\ast)+\Delta \hat{S}^{(1, 0)}_{\ell}(\Te^\ast, \Teb^\ast) (\Te/\Teb^\ast-1)
\nonumber\\
&\qquad\qquad+\Delta \hat{S}^{(2, 0)}_{\ell}(\Te^\ast, \Teb^\ast) (\Te/\Teb^\ast-1)^2,
\end{align}
where $\Delta \hat{S}^{(k,m)}_{\ell}(\Te, \Te')=(k!m!)^{-1}\Te^k\partial^k_{\Te} (\Te')^m\partial^m_{\Te'}\Delta \hat{S}_{\ell}(\Te, \Te')$. This finally gives
\beal
\label{eq:rewrite_av_two}
&\left<\sum_{\ell=0}\left[\tau_{\ell}
+[\tau\Theta]_{\ell} \Teb^\ast\partial_{\Teb^\ast} 
+\frac{1}{2} [\tau\omega^{(1)}]_\ell (\Teb^\ast)^2\partial^2_{\Teb^\ast} \right]\,\Delta \hat{S}_{\ell}(\Te, \Teb^\ast)\right>\\
&\quad\approx 
\sum_{\ell=0}\left[\left<\tau_{\ell}\right>\Delta \hat{S}^{(0,0)}_{\ell}
+\left<[\tau\Theta]_{\ell} \right>\Delta \hat{S}^{(0,1)}_{\ell}
+\left<[\tau\omega^{(1)}]_\ell\right>\Delta \hat{S}^{(0,2)}_{\ell} \right]
\nonumber\\
&\qquad
+\sum_{\ell=0}\left[\left<\tau_{\ell}\tilde{\Theta}\right>\Delta \hat{S}^{(1,0)}_{\ell}
+\left<[\tau\Theta]_{\ell} \tilde{\Theta} \right>\Delta \hat{S}^{(1,1)}_{\ell}
+\left<\tau_{\ell}\tilde{\Theta}^2\right>\Delta \hat{S}^{(2,0)}_{\ell}\right].
\nonumber
\end{align}
We suppressed the arguments of $\Delta \hat{S}^{(k,m)}_{\ell}\equiv \Delta \hat{S}^{(k,m)}_{\ell}(\Teb^\ast,\Teb^\ast)$ and defined $\tilde{\Theta}=\Te/\Teb^\ast-1$. As this expression shows, for photons scattering into the line of sight also first derivative terms of $\Delta \hat{S}_{\ell}$ contribute. 
%
%
Like for the isothermal case the computation of the spectral redistribution part, i.e. the integrals $\Delta \hat{S}_{\ell}(\Te^\ast, \Teb^\ast)$ and its derivatives, can be carried out independent of the line of sight averages over the ICM. Also, the coefficients $[\tau\Theta]_{\ell}$ and $[\tau\omega^{(1)}]_\ell$ can be computed directly with $\tau_\ell$, $\Theta_\ell$ and $\omega^{(1)}_\ell$ using Wigner-3$j$ symbols. 
This shows that the second scattering SZ signal in principle depends on spatial variations of the electron temperature.

\section{Lowest order kinematic corrections}
\label{sec:kin_corrs_lowest_order}
With our method it is also straightforward to compute the second scattering correction to the kSZ effect at the lowest order in the cluster velocity $\betac={\mathrm v_{\rm c}}/c$. 
For cold electrons, the singly scattered SZ signal is given by \citep{Sunyaev1980}
\beal
\label{eq:kSZ_formula}
\Delta I/I_{\rm o} \approx \tau \beta_\parallel \frac{x^4 \expf{x}}{(\expf{x}-1)^2}=\tau \beta_\parallel x^3 G(x),
\end{align}
where $\beta_\parallel=\vgh\cdot\vbetac=\betac \muc$ is the line of sight component of the cluster's peculiar motion, $\vbetac=\vek{\rm v}_{\rm c}/c$. The second scattering signal has two main contributions. One is of the order of $\simeq \tau^2 \betac$ and just derives from the Thomson scattering correction to the kSZ signal. The other is $\simeq \tau^2 \The \betac$, which is caused by two effects: (i) temperature-dependent corrections to the kSZ in the second scattering and (ii)  temperature-dependent corrections to the first scattering with subsequent Thomson corrections in the second scattering. The relevant temperature correction to the singly scattered signal is (see CNSN)
\beal
\label{eq:kSZ_formula_The_single}
\Delta I/I_{\rm o} &\approx \tau \beta_\parallel \The x^3 \left(\frac{2}{5}\left[G(x)+ D^{(0)}_0(x)\right]-x \partial_x Y_0 \right)\equiv \tau \beta_\parallel \The x^3 D_0,
\\ \nonumber
D^{(0)}_0(x)
&=x^{-2}\partial_x x^4 \partial_x G(x) 
\equiv - [ 4 D_x + 6 D^2_x + D^3_x ] \nbb(x).
\end{align}
The first two contributions in parenthesis arise from the motion-induced dipolar part of the radiation field, while the second term is just from the Lorentz transformation of the thSZ signal from the cluster rest frame back into the CMB rest frame.

\begin{figure}
\centering
\includegraphics[width=\columnwidth]{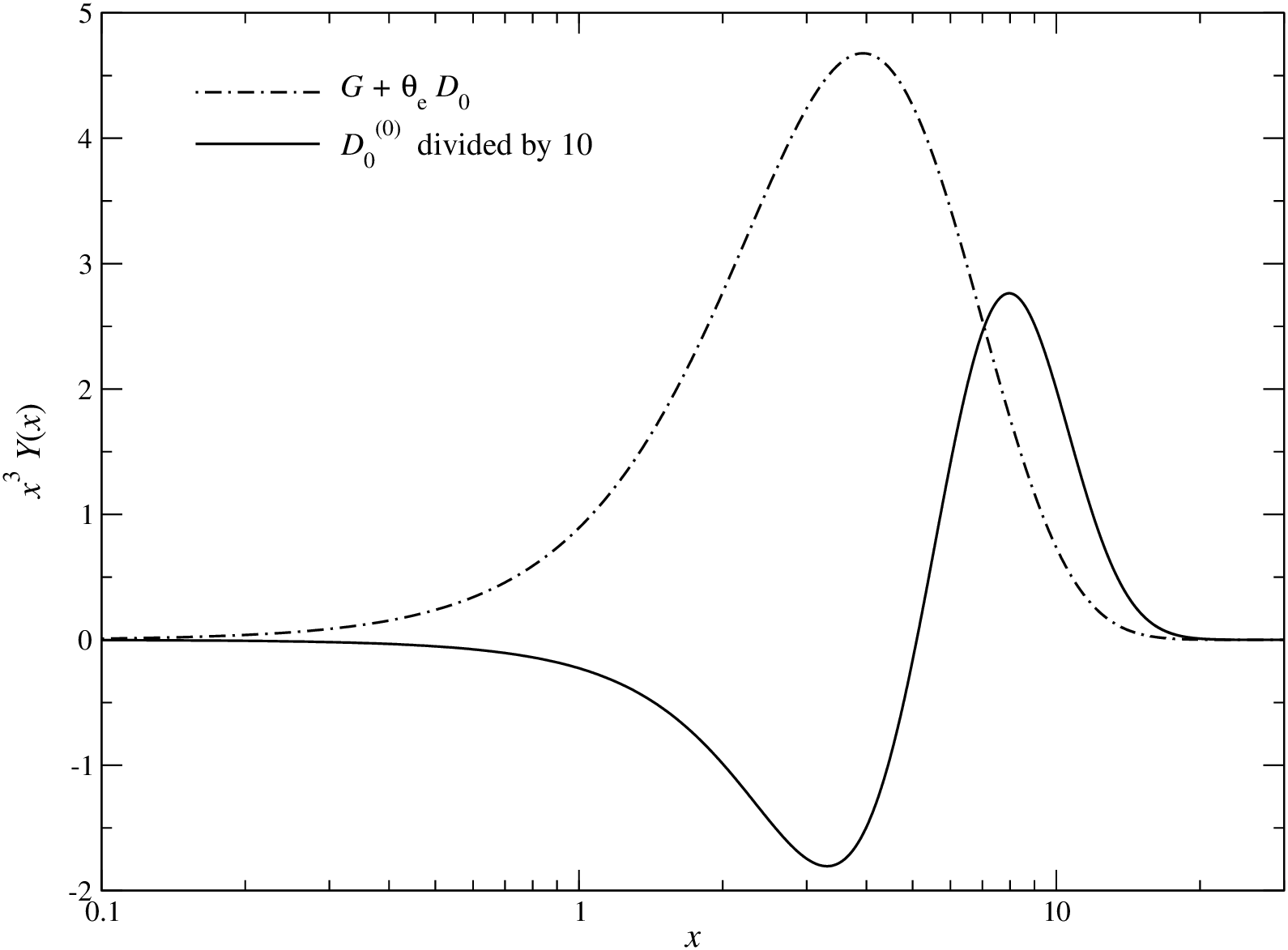}
\caption{Spectral shape of the lowest order first and second scattering contributions to the kSZ signal for $k\Te=5\,\keV$. Note that we renormalized $D^{(0)}_0=x^{-2}\partial_x x^4 \partial_x G$ to make it more comparable in amplitude to $G+\The D_0$.}
\label{fig:lowest_order_Y_functions_kSZ}
\end{figure}
To compute the second scattering correction we again have to describe the anisotropy of the singly scattered radiation field around different locations inside the cluster. For the kinematic terms now also the variation of $\beta_\parallel$ with respect to the line of sight matters. Because 
$\beta_\parallel(\vghp)=(4\pi/3) \, \betac \sum Y^\ast_{1m}(\vbetach)  Y_{1m}(\vghp)$,
the azimuthally symmetric projections with respect to $\vgh$ are
\beal
\label{eq:beta_vghp}
\left[\tau(\vek{r}, \vghp)\beta_\parallel(\vghp)\right]_{\ell}&\equiv\sqrt{\frac{2\ell+1}{4\pi}}\int Y_{\ell 0}(\vghp) \,\tau(\vek{r}, \vghp)\, \beta_\parallel(\vghp) \id^2 \vghp
\\[1mm]
&
=\frac{2\ell+1}{\sqrt{3}} \, \betac \sum_{\ell'} \sum_{m=-1}^1 \sqrt{2\ell'+1}\,\tau_{\ell' m}(\vek{r})\, Y_{1m}(\vbetach) 
\nonumber\\
&\qquad\quad \times
(-1)^m \left(\!\!
\begin{array}{ccc}
\ell  & \ell' & 1\\
0& 0& 0
\end{array}
\!\!\right)
\left(\!\!
\begin{array}{ccc}
\ell  & \ell' & 1\\\nonumber
0 & m & -m
\end{array}
\!\!\right).
\end{align}
As this expression shows, the second scattering signal in principle is sensitive to components of the optical depth field with $m=-1, 0,1$. This is because the cluster's motion introduces a preferred direction that breaks the symmetry. We can also see that the signal depends on {\it all three} components of the cluster's peculiar velocity, so that it could in principle allow constraining the large-scale cosmological velocity field. 
Including only the lowest order temperature terms with $\alpha_0=1$, $\alpha_0=-2/5$, $\alpha_2=1/10$ and $\alpha_0=-3/70$, we find the second scattering signal
\beal
\label{eq:second_lowest_DI_kSZ}
\Delta I^{(2)}(x, \vgh)&\approx \Delta I^{(2), \rm T}(x, \vgh)+\Delta I^{(2), \rm E}(x, \vgh),
\nonumber\\[1mm]
\Delta I^{(2), \rm T}/I_{\rm o}
&\approx  x^3 \left[G(x)+\The D_0\right] 
\left[ \left<[\tau\beta_\parallel]_0\right> + \frac{\left<[\tau\beta_\parallel]_2\right>}{10} 
- \frac{\tau^2 \beta_\parallel}{2} \right],
\nonumber\\[1mm]
\Delta I^{(2), \rm E}/I_{\rm o}
&\approx
\The x^3 D^{(0)}_0(x)  \sum_\ell \alpha_\ell \left<[\tau\beta_\parallel]_\ell\right>,
\end{align}
where we neglected the correction caused by the temperature-dependence of the cross section, $\Delta I^{(2), \sigma}$.
For the first and second scattering kSZ effect, the relevant spectral functions are $G(x)+\The D_0$ and $D^{(0)}_0(x)$. These are shown in Fig.~\ref{fig:lowest_order_Y_functions_kSZ}. The dominant correction arises again  in the Thomson limit; however, the second scattering kSZ correction is usually smaller than the second scattering thSZ correction, simply because $\beta_\parallel < \The$. 

In terms of geometric dependences a similar discussion as for the thSZ correction applies. In the ISA, the correction is again independent of geometry and given by $\Delta I^{(2), \rm iso}(x, \vgh)\approx \The \beta_\parallel  x^3 D^{(0)}_0(x)\, \tau^2/2$.
Including the effect of anisotropic scattering makes the situation richer. For spherical symmetry, as shown in Appendix~\ref{sec:iso_beta_taubeta}, at impact parameter $b$ we have 
\beal
\left<\left[\tau\beta_\parallel \right]_{\ell}\right>
&=
\lambda_\ell(b)\,\beta_\parallel + \kappa_\ell(b)\,\beta_\perp \cos \Delta \varphi,
\end{align}
where $\beta_\perp=\betac \sqrt{1-\muc^2}$ and $\Delta \varphi=\varphi_{\rm c}-\varphi_{\rm r}$ define the phase between the projections of $\vek{r}$ and the velocity vector $\vbetac$ on to the sky. The functions $\lambda_\ell(b)$ and $\kappa_\ell(b)$ [see Eq.~\eqref{eq:beta_vghp_rewrite}] encode the dependence on the geometry of the scattering medium. 
\begin{figure}
\centering
\includegraphics[width=\columnwidth]{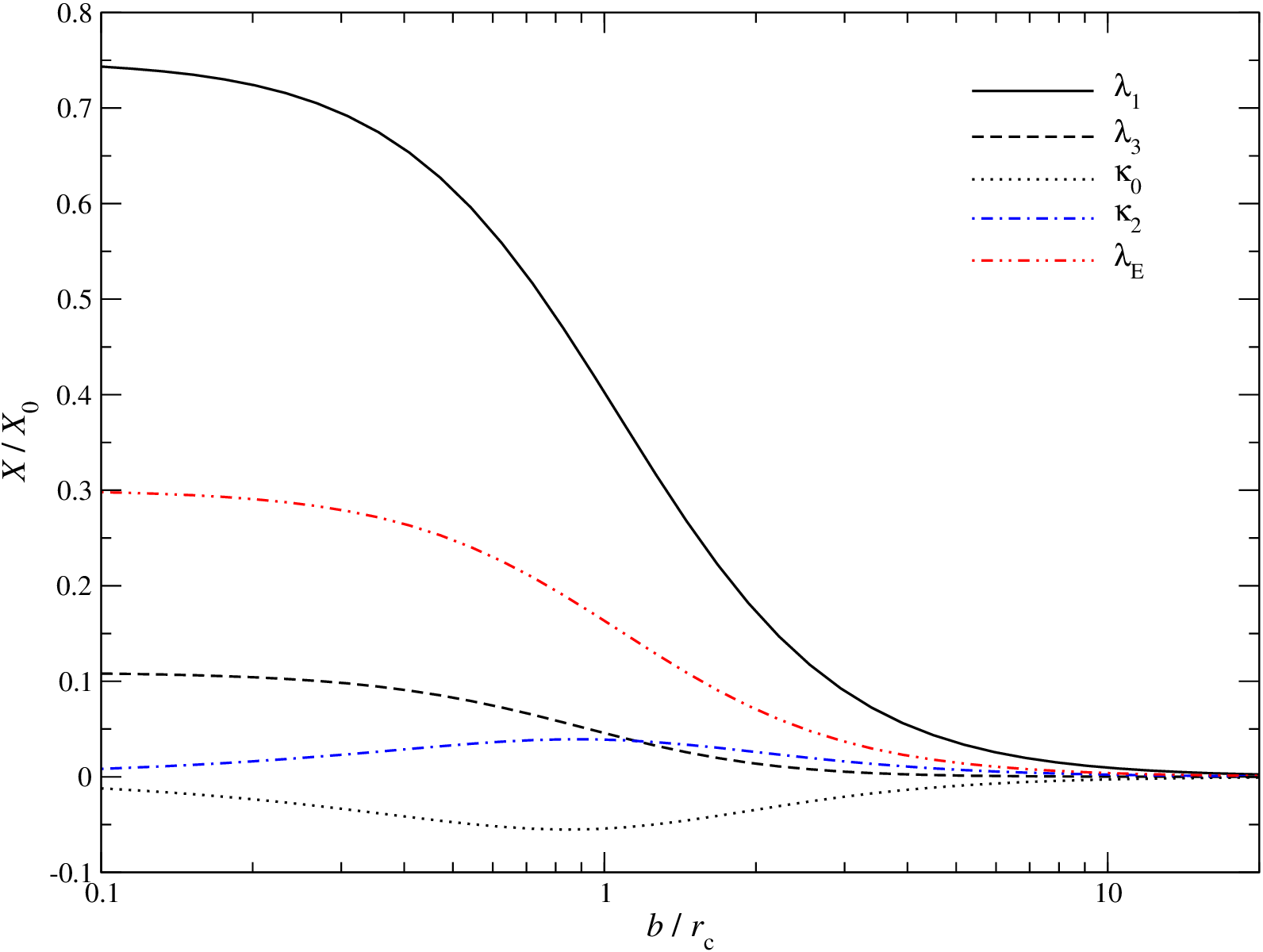}
\caption{Dependence of $\lambda_\ell$ and $\kappa_\ell$ for an isothermal $\beta$-model on the impact parameter $b$. We normalized all curves by $X_0=\tau_{\rm c}^2/2$. We separately show $\lambda_{\rm E}=(2/5)\,\lambda_1 + (3/70)\lambda_3$, which is relevant for the energy exchange term from kinematic corrections.
}
\label{fig:spatial_kin}
\end{figure}
In Fig.~\ref{fig:spatial_kin}, we illustrate their dependence on the impact parameter for an isothermal $\beta$-profile. By symmetry $\lambda_{2k}=\kappa_{2k+1}=0$. Furthermore, we find $\kappa_0\approx -\kappa_2$, because in both cases the dominant moment is $\left<\tau_{1,1}\right>$. In terms of SZ observables it is useful to group terms $\propto \beta_\parallel$ and $\propto \beta_\perp$:
\beal
\label{eq:T_E_coeffies}
\kappa_{\rm T}^{\rm kin}
&=\left<[\tau\beta_\parallel]_0\right> + \frac{\left<[\tau\beta_\parallel]_2\right>}{10} 
\approx  -\frac{\tau^2 \beta_\parallel}{2} + \frac{9}{10}\,\beta_\perp \cos \Delta \varphi \, \kappa_0,
 \\ \nonumber
\kappa_{\rm E}^{\rm kin}
&=\sum_\ell \alpha_\ell \left<[\tau\beta_\parallel]_\ell\right>
\approx -\beta_\parallel \left[ \frac{2}{5}\,\lambda_1 + \frac{3}{70}\lambda_3 \right]
+ \frac{9}{10}\,\beta_\perp \cos \Delta \varphi \, \kappa_0,
\end{align}
which shows that for the considered case three functions are relevant. The only new one is $\lambda_{\rm E}=(2/5)\,\lambda_1 + (3/70)\lambda_3$, which we  present in Fig.~\ref{fig:spatial_kin}.
 Although the coefficients $\kappa_{\rm T}^{\rm kin}$ and $\kappa_{\rm E}^{\rm kin}$ depend on {\it all} components of the cluster's peculiar velocity, the dependence on $\beta_\perp$ is very weak. Assuming that $\beta_\perp=0$, we find a Thomson scattering correction $\simeq \tau_{\rm c}/2$ relative to the kSZ signal, which could reach $\simeq 1\%$ for dense systems. The second scattering distortion, $\propto \beta_\parallel \The \lambda_{\rm E}$, only reaches $\simeq 5\times 0.3 \, \The\, \tau_{\rm c}/2 \simeq 10^{-4}$, where the factor of 5 is the rough amplitude ratio of the spectral functions $G(x)$ and $D^{(0)}_0(x)$ (cf. Fig.~\ref{fig:lowest_order_Y_functions_kSZ}). 
Next assuming $\beta_\parallel\approx \beta_\perp$ and $\Delta \phi=0$ we have the maximal effect related to $\beta_\perp$ at impact parameter $b\simeq r_{\rm c}$, with $\kappa_0\simeq -0.06\,\tau_{\rm c}^2/2$. This means that the Thomson scattering correction for $\tau_{\rm c}\simeq 0.01$ is $\simeq -(9/10)\,0.06\,\tau_{\rm c}/2\simeq \pot{2.7}{-4}$ relative to the kSZ effect. Clearly, this effect is negligibly small, and the energy exchange correction $\propto \beta_\perp$ is even smaller.

\section{Scattering of primordial CMB temperature anisotropies}
\label{sec:prim_CMB}
For the derivation of the SZ effect, the usual assumption is that the unscattered radiation field is isotropic. However, one additional, small correction to the SZ signal is caused by the scattering of primordial CMB temperature anisotropies. The incoming radiation field is given by $n^{(0)}(x, \vgh) \approx \nbb(x)+\mathcal{G}(x)\,\Theta_\gamma(\vgh)$ with $\mathcal{G}(x)=-x\,\partial_x \nbb(x)=x\,\expf{x}/[\expf{x}-1]^2$ and $\Theta_\gamma(\vgh)=\Tg(\vgh)/T_0-1$, where $\Theta_\gamma(\vgh)$ describes the CMB temperature anisotropies. The correction to the SZ signal is then given by 
\beal
\label{eq:DI_CMB_single}
\Delta I^{(1)}_{\rm prim}&=\tau \, x^3 I_{\rm o}\,\sum_{\ell=1} \Theta_{\gamma, \ell}
\int_{-\infty}^\infty \mathcal{P}_\ell(s, \Te) \, \left[ \mathcal{G}(x \expf{s})-\mathcal{G}(x)\right] \id s
\nonumber\\
&\qquad + \tau \, x^3 I_{\rm o}\,\mathcal{G}(x) \sum_{\ell=1} \Theta_{\gamma, \ell}
\left[ \int_{-\infty}^\infty \mathcal{P}_\ell(s, \Te) \,  -1\right].
\end{align}
Again two contributions appear, one that alters the spectrum of the CMB anisotropy by upscattering photons (first integral), and a second that just redistributes photons into different directions. Neglecting small temperature corrections to the total scattering cross sections of different multipoles, the latter is approximately given by \citep[see also][]{Zeldovich1980b, Sunyaev1981} $\Delta I^{(1), \rm T}_{\rm prim}(\vgh)\approx x^3 I_{\rm o}\,\mathcal{G}(x) \, \tau(\vgh) \left[\Theta_{\gamma, 2}/10 - \Theta_{\gamma}(\vgh)\right]$. 
This shows that clusters affect the CMB anisotropies like an {\it optical depth screen}, on average removing photons from the line of sight, modulating the background anisotropy in the direction of the cluster. In addition, a small part of the photons from the local quadrupole are scattered back into the line of sight.
This effect causes power from large-scale CMB isotropies ($\ell\lesssim 500$) to leak towards small scales ($\ell\simeq 2000-3000$), comparable to the angular scales of typical optical depth pockets (and troughs) hosted by clusters and other large-scale inhomogeneities \citep[for similar discussion see][]{Carlos2010}. 

A similar effect arises due to the patchiness of reionization \citep{Natarajan2012}, however, the late effects from clusters and voids as sources of optical depth variations were not discussed there. It is straightforward to estimate this contribution. Using the thSZ power spectrum \citep[e.g.,][]{Trac2011} and assuming that clusters are isothermal at average temperature $\left<k\Te\right>\simeq 5\,\keV$, one can give a template for the $\tau$-screen caused by clusters, $C^{\tau}_\ell \simeq (\left<k\Te\right>/\me c^2)^{-2} C^{\rm thSZ}_\ell \approx 10^4 C^{\rm thSZ}_\ell(\nu=148\,\rm GHz)$. For the $\tau$SZ signal mostly large-scale CMB anisotropies ($\ell \lesssim 500$) matter, so that
\beal
\label{eq:C_tau_SZ}
C^{\tau{\rm SZ}}_\ell&\approx  C^{\tau}_\ell\, \sum_{\ell=1} \frac{(2\ell+1)}{4\pi} C_\ell^{TT}  \approx \pot{2}{-5} C^{\rm thSZ}_\ell(\nu=148\,\rm GHz).
\end{align}
This is $\simeq 100$ times smaller than the estimated $\tau$-signal from patchy reionization \citep{Natarajan2012}, with an apparent root means square, $\tau_{\rm rms}\simeq 10^{-4}$ from clusters.

A few comments are in place here. Cross-correlations of the $\tau$SZ signal with the thSZ might allow identification of this contribution. It should also contribute to the SZ bispectrum \citep{Bhattacharya2012}. In addition, voids, which constitute optical depth valleys, should show up as optical depth screen, although their overall thSZ signal ought to be tiny. Given the very different characteristic scale of voids their optical depth screening effect might be separable from the $\tau$SZ and patchy reionization $\tau$ signal.
Also, since the spectrum of this $\tau$SZ signal is thermal, it is fully degenerate with the cluster's kSZ effect; however, the possible bias is no larger than a few percent of the kSZ signal, unless there is an unusually large primordial dipole, $\Theta_{\gamma, 1}\simeq 10^{-4}-10^{-3}$. 
And finally, because we can directly measure the large-scale CMB anisotropies, knowing the positions of thSZ clusters we can in principle predict the $\tau$SZ signal for our realization of the Universe. This avoids limitations set by cosmic variance of the large scales CMB modes.
Still, the overall effect is very small, and hence unobservable at this stage.

\begin{figure}
\centering
\includegraphics[width=\columnwidth]{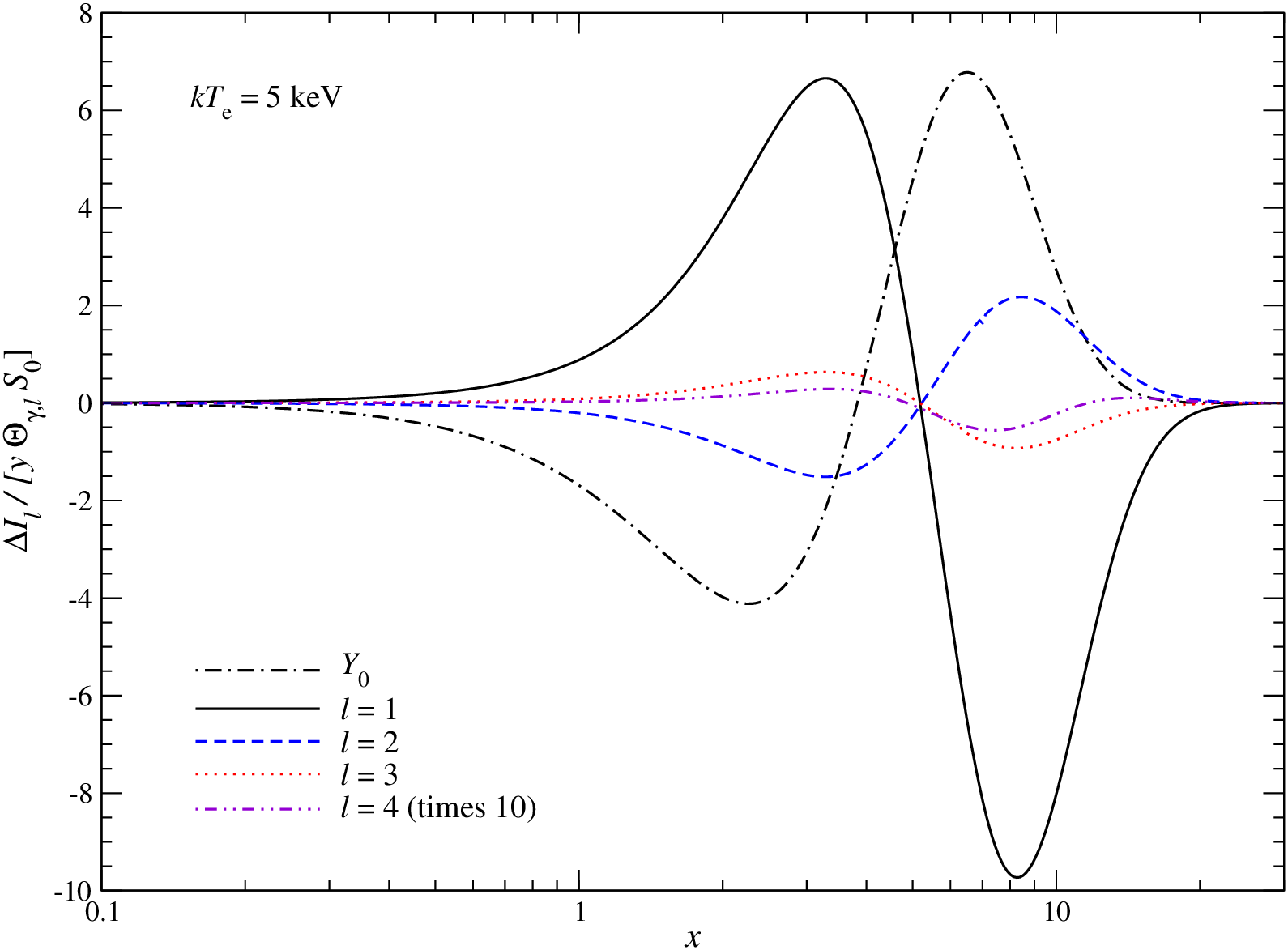}
\\[2mm]
\includegraphics[width=\columnwidth]{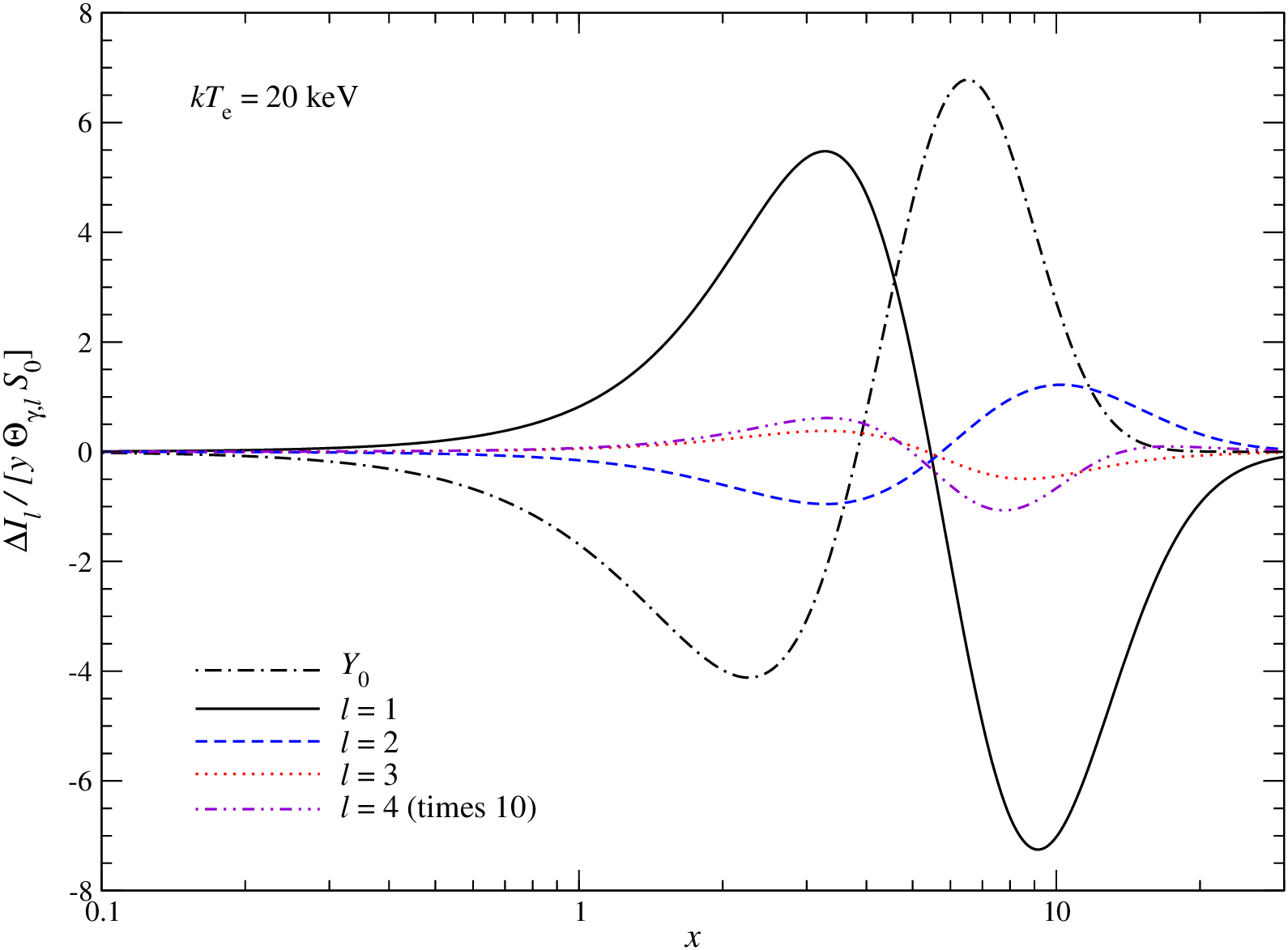}
\caption{Scattering of CMB temperature anisotropies by SZ clusters of different temperature. For reference we also show the spectral shape for the thSZ signal. Note, however, that the signal caused by CMB temperature anisotropies is suppressed by $\simeq \Theta_{\gamma, l}$.}
\label{fig:CMB_anisotropy_signals}
\end{figure}
The distortion part of $\Delta I^{(1)}_{\rm prim}$ again can be computed without direct reference to the actual amplitude of the CMB temperature anisotropies or the cluster's optical depth. In Fig.~\ref{fig:CMB_anisotropy_signals}, we show the spectral behavior for the first few multipoles. The signal is at least $\simeq \Theta_{\gamma, \ell}$ times smaller than the thSZ of the same cluster. It is largest for scattering of the local dipole anisotropy and drops very strongly for $\ell>3$, due to additional temperature suppression. The signal is therefore negligible overall. This can be also shown by considering the exact position of the SZ crossover frequency:
at the lowest order in $\The = k\Te/\me c^2$, we have 
\beal
\label{eq:DI_CMB_single_E}
\Delta I^{(1), \rm E}_{\rm prim}
&= 
\tau \, x^3 I_{\rm o}\,\sum_{\ell=1} \Theta_{\gamma, \ell}
\int_{-\infty}^\infty \mathcal{P}_\ell(s, \Te) \, \left[ \mathcal{G}(x \expf{s})-\mathcal{G}(x)\right] \id s
\nonumber\\
&\approx \tau \, x^3 \The I_{\rm o}\, \sum_{\ell=1}^{3} \alpha_\ell \,\Theta_{\gamma, \ell} \, D^{(0)}_0(x) ,
\end{align}
where $D^{(0)}_0(x)$ is given by Eq.~\eqref{eq:kSZ_formula_The_single} and shown in Fig.~\ref{fig:lowest_order_Y_functions_kSZ}. Thus, the SZ null is shifted by 
$\Delta \xc \approx 0.38  [ -4 \Theta_{\gamma, 1} + \Theta_{\gamma, 2} - 0.4 \Theta_{\gamma, 3}]$,
which gives no more than $|\Delta \xc | \simeq  \pot{\rm few}{-5}$. Nevertheless, this contribution can be easily accounted for once the position of the cluster relative to the CMB anisotropies (dipole through quadrupole) is known. In particular, a large primordial CMB dipole should leave a signature with known angular dependence. This could provide a way to place upper limits on the amplitude of the primordial dipole using large SZ cluster samples, an effect that is complementary to the SZ cluster number count anisotropy caused by the motion of the Solar system with respect to the CMB \citep{Chluba2005b}.

\section{Conclusion}
\label{sec:conclusions}
We demonstrated that the second scattering SZ signal at high temperatures has non-trivial dependence on the spatial variations of the electron temperature and density (see Fig.~\ref{fig:SZ_signal_multi-scatt.Te_5keV}). These aspects are not captured by the ISA used in previous analysis of the problem.
A detection of these differences will be challenging, especially when including spatially varying foregrounds and instrumental effect; however, they give rise to another independent set of SZ observables that in the future could be important for the 3D cluster-profile reconstruction.
Also, due to the $\simeq \Ne^2$ scaling, the detectability of the signals could be further increased by using cross-correlation with X-rays (CDK13), but a more detailed assessment of this possibility is beyond the scope of this work.

Including variations of the electron temperature along different lines-of-sight in principle further increases the total number of independent SZ variables, owing to the fact that different contributions all have different spectra [cf. Eq.~\eqref{eq:non_iso_DI2}]. 
Because the models used here to demonstrate the effects are very simplistic, it would be interesting to quantify the dependences in more detail using realistic cluster simulations. In this way, it is also possible to account for the full asphericity and clumpiness of the ICM, which could increase the importance of the second scattering SZ effect.

We showed that the lowest order kinematic corrections depend on all three components of the cluster's peculiar motion [see Eq.~\eqref{eq:second_lowest_DI_kSZ}]. While this is intriguing from an observational point of view, the overall signal again is extremely small, rendering an application of this effect challenging (see Sect.~\ref{sec:kin_corrs_lowest_order}).

In Sect.~\ref{sec:prim_CMB}, we discussed the SZ signal caused by scattering of primordial CMB isotropies. Clusters host free electrons which in addition to the thSZ and kSZ signals (from scattering of the CMB monopole) also create a $\tau$SZ effect (mainly by scattering CMB background photons out of the line of sight) in the Thomson limit. The signal is thermal and appears as contribution to the small-scale CMB temperature power spectrum. However, our estimates show that the signal is about 100 times smaller than, e.g., the signal caused by optical depth variation during the reionization epoch \citep{Natarajan2012}.
Additional variation of the average optical depth could come from voids, which were not discussed here but could enhance the overall effect at slightly different scales. The $\tau$SZ effect could furthermore bias kSZ measurements \citep[see][for more discussion]{Carlos2010}.
Spectral distortions created by the scattering of the primordial CMB dipole through octupole by hot electrons in clusters also leave a signature in the SZ signal, but as discussed in Sect.~\ref{sec:prim_CMB}, this effect is small. 
While a detection of the $\tau$SZ effect will be challenging, in the future the positions of millions of SZ clusters will be known from thSZ measurements. This could provide an approximate spatial template for the $\tau$SZ contribution, at least for the largest system, which again could be used to enhance the observability of this small effect; however, a more detailed discussion of future observational prospects is left to another work.

\section*{Acknowledgements}
The authors thank Marc Kamionkowski for stimulating discussions and suggestions. They are also grateful to Rashid Sunyaev for comments on the manuscript. JC furthermore thanks Nick Battaglia for useful discussion on the $\tau$SZ effect. This work is supported by the grants DoE SC-0008108 and NASA NNX12AE86G.

\begin{appendix}

\small

\section{Boltzmann collision term for Compton scattering}
\label{app:collision_term}

\subsection{Photon Boltzmann equation and collision term}
\label{sec:Boltz_Coll_photons}
For hot electrons, $\Te \gg T_0$, the Boltzmann collision term is explicitly discussed in Sect.~2.1 of CNSN. 
Aligning the $z$-axis with $\vgh$, and writing $n(x, \vek{r}, \vgh)$ as spherical harmonic expansion, one obtains
\beal
\label{eq:Boltzmann_expansion}
\mathcal{C}[n]
&\approx  
\int \frac{f_{\rm e}(p)}{\Ne \sigT} \frac{\id \sigma}{\id \Omega'} 
\left[ n(x', \vek{r}, \vghp) - n(x, \vek{r}, \vgh)\right] \!\id^2\vghp\!\id^3 p
\nonumber\\
&=
\sum_{\ell,m}  \int \frac{f_{\rm e}(p)}{\Ne \sigT} \frac{\id \sigma}{\id \Omega'} \,
n_{\ell m}(x', \vek{r}) \, Y_{\ell m}(\vghp) \id^2\vghp\!\id^3 p  - n(x, \vek{r}, \vgh)
\nonumber\\
&\equiv 
\sum_\ell  \int \frac{4\pi f_{\rm e}(p)}{\Ne \sigT} \frac{\id^2 \sigma_\ell}{\id \mu \!\id \mu'} \,
n_{\ell}(x', \vek{r}) \, p^2 \!\id p \id \mu' \!\id\mu - n(x, \vek{r}, \vgh),
\nonumber\\
\frac{\id^2 \sigma_\ell}{\id \mu \!\id \mu'}
&=
\!\int P_{\ell}(\mu_{\rm sc})\, \frac{\id \sigma}{\id \Omega'} 
\frac{\id\varphi\id \varphi'}{4\pi},
\end{align}
where\footnote{For a general coordinate system one has $n_{\ell} \equiv \sum_{m=-\ell}^\ell n'_{\ell m}Y_{\ell m}(\vgh)$.} $n_{\ell}=\sqrt{(2\ell+1)/(4\pi)} \,n_{\ell 0}$ and $P_\ell(x)$ is a Legendre polynomial; $f_{\rm e}(p)$ is the relativistic Maxwell-Boltzmann distribution of electrons (Eq.~(4) of CNSN); $\id \sigma/\id \Omega'$ is the differential cross section for Compton scattering (Eq.~(2) of CNSN); and the direction cosines are $\mu=\vbh\cdot \vgh$, $\mu'=\vbh\cdot \vgh'$ and $\mu_{\rm sc}=\vgh\cdot \vghp$. Here, $\vbh$ and $\vghp$ define the direction of the incoming electron and photon, respectively.
Finally, $x'\approx x (1-\beta\mu)/(1-\beta\mu')$ when neglecting recoil effects, which is possible since $T_0\ll \Te$.

For the last step in the definition of $\mathcal{C}[n]$, we used the symmetry of the scattering process. After integrating over $\varphi$, the cross section becomes independent of $\varphi_{\rm sc}$ (also see CNSN). Therefore, only multipoles of the radiation field with $m\equiv 0$ matter for the scattering of photons into the line of sight, and it is sufficient to only keep terms $\propto n_{\ell 0}\, Y_{\ell 0} \propto P_\ell(\mu_{\rm sc})$. With this simplification, it is best to use $\id\varphi'\id\mu'$ instead of $\id\varphi_{\rm sc}\id\mu_{\rm sc}$ to carry out the integral over the incoming photon distribution, applying the identity $\mu_{\rm sc}=\mu\mu'+\cos(\varphi-\varphi')\sqrt{1-\mu^2}\sqrt{1-\mu'^2}$.
Explicit expressions for $\id^2 \sigma_\ell/(\id \mu \!\id \mu')$ with $\ell\leq 2$ are given in Appendix~B of CNSN.
Using the addition theorem for spherical harmonics, one can find the general expression for $\ell>0$
\beal
\label{eq:sig_av_gen}
\frac{\id^2 \sigma_\ell}{\id \mu \!\id \mu'}
&=
P_\ell(\mu)P_\ell(\mu')\frac{\id^2 \sigma_0}{\id \mu \!\id \mu'}
\\
&+
\frac{3\sigT}{8}\,\zeta^2\gamma^2\kappa^3
\left\{
\zeta(1-\zeta)\frac{(\ell-1)!}{(\ell+1)!} 
P^1_\ell(\mu)P^1_\ell(\mu')\,P^{1}_1(\mu)P^{1}_1(\mu')
\right.
\nonumber\\
&\qquad+
\left.
\frac{2}{3}\zeta^2\sum_{m=1}^{2}\frac{(\ell-m)!(2-m)!}{(\ell+m)!(2+m)!} 
P^m_\ell(\mu)P^m_\ell(\mu')\,P^{m}_2(\mu)P^{m}_2(\mu')
\right\},
\nonumber
\end{align}
where we defined $\zeta=\nu'/(\nu\gamma^2\kappa^2)$ and $\kappa=1-\beta \mu$. We note that this result is only valid if recoil and stimulated terms can be neglected ($T_0/\Te\ll1$).

\subsubsection{Fokker-Planck expansion of the collision integral}
\label{sec:collision_term_moments}
To compute the collision integral over incoming photons, one can formally rewrite the photon distribution performing a Taylor expansion in the frequency shift, $\Delta_\nu=(\nu'-\nu)/\nu$. Defining the moments of the scattering kernel
\beal
\label{eq:moment_kernel}
I^{k}_{\ell} \equiv \left<\Delta_\nu^k\right>_\ell
&=  
\int \frac{4\pi f_{\rm e}(p)}{\Ne \sigT} \frac{\id^2 \sigma_\ell}{\id \mu \!\id \mu'} \,\frac{\Delta_\nu^k}{k!}
\, p^2 \!\id p \id \mu' \!\id\mu,
\end{align}
we can cast Eq.~\eqref{eq:Boltzmann_expansion} into the form
\beal
\label{eq:dist_second_rewrite}
\mathcal{C}[n]
&\approx \sum_{\ell=0}  \sum_{k=0} I^{k}_\ell \, x^{k}\partial^k_x n_{\ell}(x, \vek{r}) - n(x, \vek{r}, \vgh).
\end{align}
The monopole spectrum is not affected by scattering, unless energy transfer is included ($k>0$). One can also distinguish terms without energy transfer ($k=0$) from those with energy transfer ($k>0$). The former leave the incoming photon spectrum unaltered, but help isotropizing the photon field contributing a temperature-dependent shear viscosity, while the latter also cause redistribution of photons over frequency, and hence create a distortion. 

Although formally valid, it is also well known that the collision term in the form Eq.~\eqref{eq:dist_second_rewrite} converges very slowly once the temperature of the medium becomes larger than $k\Te \simeq 10\,\keV$ \citep[e.g., see][]{Challinor1998, Sazonov1998}. Still for analytical considerations this form is very useful and applicable to general incoming radiation fields.
At the lowest order of the electron temperature, the kinetic equation describing the scattering of monopole through octupole anisotropy was previously derived and discussed by \citet{Chluba2012}. Here, we provide analytic forms of the higher order temperature corrections and explicitly compute the collision term numerically.

\subsubsection{Collision integral in the kernel approach}
\label{sec:collision_term_kernel}
The collision integral can be reformulated using the logarithmic frequency shift, $s=\ln(\nu'/\nu)$ \citep[e.g., see][]{Wright1979}. Substituting $\mu(s, \beta)=[1-\expf{s} (1-\beta \mu')]/\beta$, we can convert the $\mu$-integral into an integral over $s$. The integration range of $s$ then depends on $\mu'$, i.e. $(1-\beta)/(1-\beta\mu')\leq \expf{s} \leq (1+\beta)/(1-\beta\mu')$. Next, one can switch the $s$- and $\mu'$-integrals, so that  afterwards $-s_{\rm lim}\leq s \leq s_{\rm lim}$ and $\max[-1, \mu'_1(s, \beta)]\leq \mu' \leq \min[\mu'_2(s, \beta), 1]$ with 
\beal
s_{\rm lim}=\ln[(1+\beta)/(1-\beta)],
\nonumber\\
\mu'_1(s, \beta)=[1-\expf{-s}(1+\beta)]/\beta,
\nonumber\\
\mu'_2(s, \beta)=[1-\expf{-s}(1-\beta)]/\beta.
\end{align}
Using $\beta=\eta/\sqrt{1+\eta^2}$ and $\eta=p/(\me c)$, one can finally interchange the $s$- and $p$-integrals finding
\beal
\label{eq:Boltzmann_kernel}
\mathcal{C}[n]
&\approx- n(x, \vek{r}, \vgh) + \sum_\ell \int_{-\infty}^\infty \id s \, \mathcal{P}_\ell(s, \The) \,n_{\ell}(x \, \expf{s}, \vek{r}),
\\ \nonumber
\mathcal{P}_\ell(s, \The)&=
\int^\infty_{\eta_{\rm min}(s)} \int_{\mu'_1(s, \eta)}^{\mu'_2(s, \eta)}\,\frac{\varepsilon\,\eta \,\kappa \, \expf{-\varepsilon/\The}}{\The K_2(1/\The)\,\sigT} \frac{\id^2 \sigma_\ell}{\id \mu \!\id \mu'}  \id \mu' \id \eta,
\end{align}
with $\varepsilon(\eta)=\sqrt{1+\eta^2}$, $\eta_{\rm min}=\sinh( |s|/2 )$ and $\The=k\Te/[\me c^2]$, and where $K_2(x)$ is the modified Bessel-function of second kind. For $\ell=0$, this expression for the kernel is in agreement with the result of \citet{Wright1979}; however, here we used slightly different variables.
We mention that $\int \mathcal{P}_0(s, \The)\id s =1$ at all temperatures, and $\int \mathcal{P}_2(s, 0)\id s =1/10$, while $\int \mathcal{P}_\ell(s, 0)\id s=0$ otherwise. More generally, $\int \mathcal{P}_\ell(s, \The)\id s =\sigma_\ell(\The)/\sigT$, where $\sigma_\ell(\The)$ is the total scattering cross section of multipole $\ell$.

\begin{figure}
\centering
\includegraphics[width=\columnwidth]{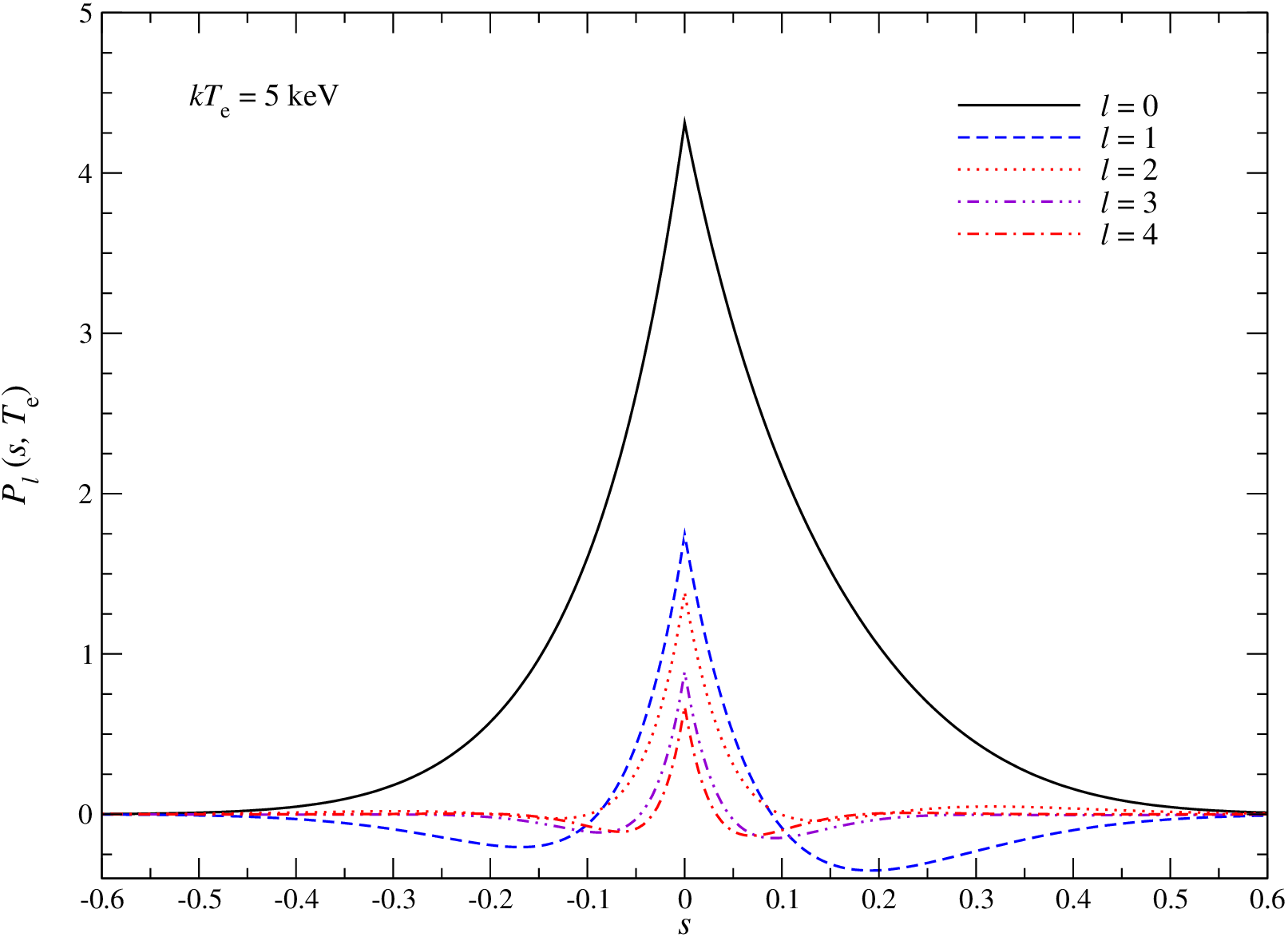}
\\[2mm]
\includegraphics[width=\columnwidth]{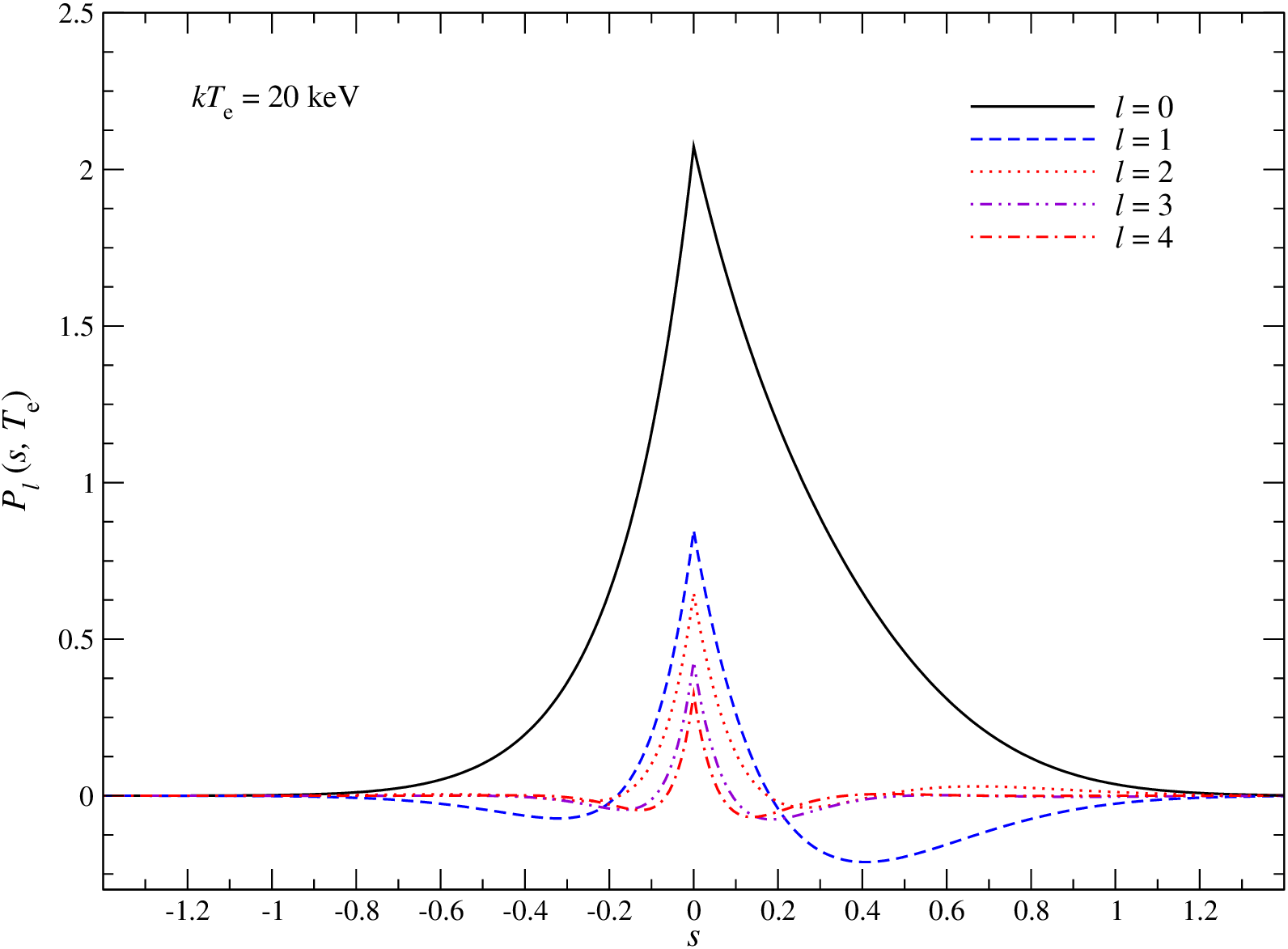}
\caption{Shape of the scattering kernel for $\ell\leq 4$. At higher temperatures the characteristic width of the scattering kernel increases (note the change in the scale of $s=\ln(\nu'/\nu)$ between the upper and the lower panel). Scattering of the monopole is strongest, while photon anisotropies scatter less efficiently as $\ell$ increases. Although the total integral over the scattering kernel decreases rapidly for $\ell>3$, the redistribution terms remain significant.}
\label{fig:Kernel}
\end{figure}

For numerical purposes, the scattering kernel formulation is very convenient, as the kernel can be precomputed for different temperatures, while the convolution of the spectrum can be carried out independently. This accelerates the numerical evaluation of the collision term by a large factor. We tabulated the kernels for $\ell\leq 4$ over a wide range of temperatures ($1\,\keV\lesssim k\Te \lesssim 100\,\keV$) for {\sc SZpack}. In particular, the second scattering correction is readily evaluated using this method, with precision $\simeq 0.1\%$ (on the correction).
In Fig.~\ref{fig:Kernel}, we show the scattering kernel for a few cases. They all exhibit a cusp at $s=0$, which is a consequence of the $s$-dependence of $\eta_{\rm min}$ and also well known for $\ell=0$ \citep[e.g.,][]{Syunyaev1980Kernel, Sazonov2000}. One additional aspect is that the kernels for $\ell>0$ are no longer positive at all frequencies. Physically, this is connected to the fact that, in a narrow frequency range, photons of the anisotropic part are redistributed not only over frequency but also spatially, leading to isotropization and damping of spectral anisotropies. However, a more detailed discussion is beyond the scope of this paper.

\section{Fokker-Planck result for second scattering terms}
\label{app:FP_result}
Using the Fokker-Planck approach, for the scattering of the isotropic CMB we can formally define $\hat{s}^{(0)}(x, \Te)=\mathcal{C}[\nbb]=\sum_{k=1} I^k_0(\Te) \,x^k\partial_x^k \nbb\equiv \The\sum_{k=0} \The^{k} Y_k(x)\equiv \The (\vek{\The}\cdot \vek{Y})$, where we introduced the vectors $\vek{\The}=(1, \The, \The^2,..., \The^n)^{\rm T}$ and $\vek{Y}= (Y_0, Y_1,..., Y_n)^{\rm T}$, with $n$ denoting the maximal order of temperature corrections (usually $n=10$). 
The vector $\vek{Y}$ itself can be determined by a simple matrix operation, $\vek{Y}=(\mathsf{M}^0\, \hat{\vek{O}})\nbb$, where the coefficients $\mathsf{M}^0_{km}\equiv a^{(k)}_m$ are defined in Table~B1 of CNSN and the operator
$\hat{\vek{O}}=(x\partial_x, x^2\partial^2_x,..., x^{2n+2}\partial^{2n+2}_x)^{\rm T}$ was introduced. Note that $\mathsf{M}$ has $n+1$ rows and $2(n+1)$ columns.
Consequently, one has $\hat{s}^{(0)}(x, \Te)=\The [\vek{\The}\cdot (\mathsf{M}^0\, \hat{\vek{O}})]\,\nbb$.

The matrix $\mathsf{M}^0$ was derived for the scattering of the monopole radiation field. Similarly, one can define the matrices $\mathsf{M}^1$, $\mathsf{M}^2$ and $\mathsf{M}^3$ using Tables~B2 and B3 from CNSN and Table~\ref{tab:oct} below to describe the effect of dipole, quadrupole and octupole scattering. For fixed multipole, we find 
\beal
\Delta I^{(2), \rm E}_\ell/[x^3 I_{\rm o}] &=\sum_{k=1} I^{k}_\ell(\Te) \, x^{k}\partial^k_x \hat{s}^{(0)}(x, \Te)
=\The [\vek{\The}\cdot (\mathsf{M}^\ell\, \hat{\vek{O}})]\,\hat{s}^{(0)}(x, \Te)
\nonumber\\
&=\The^2 [\vek{\The}\cdot (\mathsf{M}^\ell\, \hat{\vek{O}})][\vek{\The}\cdot (\mathsf{M}^0\, \hat{\vek{O}})]\,\nbb.
\end{align}
To simplify this expression at different orders of the electron temperature the operator relation
\beal
 x^{k}\partial^k_x  x^{m}\partial^m_x \nbb
 =
\sum_{j=0}^k \frac{m!}{(m-k+j)!}\,
\binom{k}{j}\,x^{m+j}\partial_x^{m+j} \nbb
\end{align}
is very useful.
Keeping temperature terms up to $\mathcal{O}(\The^{n+1})$, we find
\beal
\label{eq:result_expansion_DI2E_l_app}
\Delta I^{(2), \rm E}_\ell
&= \The^2 x^3 I_{\rm o} \sum^{n-1}_{k=0}\The^k Y^{(\ell)}_k,
\\ \nonumber
Y^{(\ell)}_k
&=\sum_{t=0}^{k} 
\sum_{i=0}^{2k+1} \sum_{j=0}^{2k+1} 
\sum_{m=0}^{i+1} 
\frac{(j+1)! \, \mathsf{M}^\ell_{k-t, i}\,\mathsf{M}^0_{t, j}}{(j-i+m)!}\,
\binom{i+1}{m}\,x^{j+1+m}\partial_x^{j+1+m} \nbb.
%
\end{align}
Although a bit clunky, the sums can be easily computed for different cases.

\subsection{Evaluation of $\left<[\tau\beta_\parallel]_{\ell}\right>$ for isothermal $\beta$-profile}
\label{sec:iso_beta_taubeta}
To illustrate the second scattering corrections to the kSZ effect, we need the moments, $\left<[\tau\beta_\parallel]_{\ell}\right>$. The simplest line of sight is the one that is aligned with $\vbetac$. In this case, $\beta_\parallel(\vghp)=\betac P_1(\vgh\cdot \vghp)=\betac\,\mu_{\rm r}$, which is independent of the position $z$ along the line of sight.
We find
\beal
\hat{[\tau\beta_\parallel]}_{\ell}(z)&=
\frac{(2\ell+1)\,\betac}{2}\int_{-1}^1 \int_0^\infty \Ne\left(\sqrt{z^2+s^2+2 \mu_{\rm r} s z}\right)\sigT \mu_{\rm r} P_\ell(\mu_{\rm r}) \id s \id \mu_{\rm r}
\nonumber\\
&\equiv
\betac[(\ell+1)\hat{\tau}_{\ell+1}(z)+\ell\,\hat{\tau}_{\ell-1}(z)],
\nonumber\\
\left<[\tau\beta_\parallel]_{\ell}\right>&=
2 \int_{0}^{\infty} \hat{[\tau\beta_\parallel]}_{\ell}(z) \, \Ne(z)\sigT \id z,
\end{align}
consistent with Eq.~\eqref{eq:beta_vghp}. This implies
\beal
\label{eq:tau_beta_para}
\left<[\tau\beta_\parallel]_{0}\right>&=(\betac/3)\left<\tau_{1}\right> \equiv 0,
\nonumber\\
\left<[\tau\beta_\parallel]_{1}\right>&= \betac[ \left<\tau_{0}\right>+(2/5)\left<\tau_{2}\right> ]=0.75 \, \betac (\tau_{\rm c}^2/2),
\nonumber\\
\left<[\tau\beta_\parallel]_{2}\right>&= \betac[ (2/3)\left<\tau_{1}\right>+(3/7)\left<\tau_{3}\right> ] =0,
\nonumber\\
\left<[\tau\beta_\parallel]_{3}\right>&= \betac[ (3/5)\left<\tau_{2}\right>+(4/9)\left<\tau_{4}\right> ]=
0.11 \, \betac (\tau_{\rm c}^2/2),
\end{align}
at impact parameter $b=0$. For the kinematic corrections only the odd moments of the optical depth-velocity field remain, while for the thSZ correction the even moments of the optical depth mattered. The case for general impact parameter $b$ is also directly obtained with Eq.~\eqref{eq:tau_beta_para} by computing the values of $\left<\tau_\ell\right>$ at $b\neq 0$.

To give the result for general line of sight and orientation of $\vbetac$, we also have to calculate the averages $\left<\tau_{\ell,-1}\right>$ and $\left<\tau_{\ell,1}\right>$, cf. Eq.~\eqref{eq:beta_vghp}. Using the addition theorem for spherical harmonics, it is straightforward to show that $\tau_{\ell m}\equiv (4\pi/[2\ell+1]) \, \tau^\ast_{\ell}(r)\,Y^\ast_{\ell m}(\hat{\vek{r}})$, where $\tau^\ast_\ell(r)$ is evaluated in the preferred frame with $z$-axis parallel to the radial vector $\vek{r}$. This means
\beal
\nonumber
\frac{\left<\tau_{\ell m}\right>}{4\pi}&=
 \sqrt{\frac{(2\ell+1)}{4\pi}\frac{(\ell-m)!}{(\ell+m)!}}\,\expf{-{\rm i}m\varphi_{\rm r}} \!\!
\int_{0}^{\infty} \!2 \tau^\ast_{\ell}(r)P^m_\ell(s/r) \, \Ne(r)\sigT \id s,
\end{align}
where $\ell+m$ has to be even and $\varphi_{\rm r}$ is the azimuthal angle of $b$ in the plane of the sky. One can set $\varphi_{\rm r}=0$, since for the SZ signal only the difference of $\varphi_{\rm r}$ and $\varphi_{\rm c}$ for the peculiar motion matters. By construction $\left<\tau_{\ell, -m}\right>=(-1)^m \left<\tau_{\ell m}\right>^\ast$, so that only the transform for $m=1$ has to be computed. Defining, $\left<\tilde{\tau}_{\ell m}\right>=\int_{0}^{\infty} 2\hat{\tau}_{\ell}(r_0)P^m_\ell(s/r_0) \, \Ne(r_0)\sigT \id s$, with Eq.~\eqref{eq:beta_vghp} after some manipulations we can write
\beal
\label{eq:beta_vghp_rewrite}
\left<\left[\tau\beta_\parallel \right]_{\ell}\right>
&=
\lambda_\ell(b)\,\beta_\parallel + \kappa_\ell(b)\,\beta_\perp \cos \Delta \varphi ,
\nonumber\\
\lambda_\ell(b) &=
\ell \left<\tilde{\tau}_{\ell-1,0}\right>+(\ell+1)\left<\tilde{\tau}_{\ell+1,0}\right>,
\nonumber\\
\kappa_\ell(b)
&=\left<\tilde{\tau}_{\ell-1, 1}\right>-\left<\tilde{\tau}_{\ell+1, 1}\right>,
\end{align}
where $\beta_\perp=\betac \sqrt{1-\muc^2}$ and $\Delta \varphi=\varphi_{\rm c}-\varphi_{\rm r}$ defines the phase between the projections of $\vek{r}$ and the velocity vector $\vbetac$ on to the sky. We also used $\left<\tau_{l}\right>=(2l+1)\left<\tilde{\tau}_{l,0}\right>$. For spherically symmetric electron distributions, $\lambda_{2k}(b)=\kappa_{2k+1}(b)=0$. Also, by symmetry $P^m_\ell(-x)=(-1)^{m+\ell}P^m_\ell(x)$ implying $\left<\tilde{\tau}_{2k, 1}\right>=0$.

\begin{table*}
\centering
\caption{Moments for octupole scattering. Blank means the coefficient is zero. In each row, the temperature order increases, while in each column the derivative of a Planckian increases. 
The columns define the functions $O_n(x)=\sum_{k=1}^{2n+2} o^{(n)}_k x^k\partial^k_x \nbb(x)$, while the rows define the moments, $I^k_\ell(\The)=\The\,\sum_{m=0}^n \The^m \, o^{(m)}_k$.}
\begin{tabular}{cccccccccc}
& $O_0$ 
& $O_1$ 
& $O_2$ 
& $O_3$ 
& $O_4$ 
& $O_5$ 
& $O_6$ 
& $O_7$ 
& $O_8$ 
\\
\hline
\hline
$I^0_3$ & $\frac{6}{35} $ 
& $ -\frac{57}{35} $ 
& $ \frac{1517}{140} $ 
& $ -\frac{1833}{28} $ 
& $ \frac{1941243}{4928} $ 
& $ -\frac{3025335}{1232} $ 
& $ \frac{8263393335}{512512} $ 
& $ -\frac{1796776893}{16016} $ 
& $ \frac{13585352311395}{16400384} $
\\[1mm]
$I^1_3$ & $-\frac{6}{35} $ 
& $ \frac{57}{35} $ 
& $ -\frac{1517}{140} $ 
& $ \frac{1833}{28} $ 
& $ -\frac{1941243}{4928} $ 
& $ \frac{3025335}{1232} $ 
& $ -\frac{8263393335}{512512} $ 
& $ \frac{1796776893}{16016} $ 
& $ -\frac{13585352311395}{16400384} $
\\[1mm]
$I^2_3$ & $ -\frac{3}{70} $ 
& $ -\frac{123}{140} $ 
& $ \frac{883}{560} $ 
& $ -\frac{4647}{112} $ 
& $ \frac{535251}{2816} $ 
& $ -\frac{6018297}{4928} $ 
& $ \frac{16520914089}{2050048} $ 
& $ -\frac{3594051855}{64064} $ 
& $ \frac{27172907535645}{65601536} $
\\[1mm]
$I^3_3$  &  
& $ -\frac{18}{35} $ 
& $ -\frac{1108}{105} $ 
& $ -\frac{2854}{35} $ 
& $ -\frac{388483}{770} $ 
& $ -\frac{97049}{154} $ 
& $ -\frac{111750291}{32032} $ 
& $ \frac{155979471}{8008} $
& $ -\frac{35476758123}{256256} $
 \\[1mm]
$I^4_3$ &  
& $ -\frac{3}{70} $ 
& $ -\frac{629}{105} $ 
& $ -\frac{10019}{70} $ 
& $ -\frac{5100451}{3080} $ 
& $ -\frac{7118137}{616} $ 
& $ -\frac{5936654643}{128128} $ 
& $ -\frac{3564673617}{32032} $ 
& $ -\frac{50713068171}{1025024} $
 \\[1mm]
$I^5_3$&  &  
& $ -\frac{92}{105} $ 
& $ -\frac{2386}{35} $ 
& $ -\frac{1369777}{770} $ 
& $ -\frac{19527023}{770} $ 
& $ -\frac{36629425317}{160160} $ 
& $ -\frac{54269171223}{40040} $ 
& $ -\frac{1335158910633}{256256} $
 \\[1mm]
$I^6_3$& &
& $ -\frac{23}{630} $ 
& $ -\frac{5113}{420} $ 
& $ -\frac{13806257}{18480} $ 
& $ -\frac{385323503}{18480} $ 
& $ -\frac{447802072119}{1281280} $
& $ -\frac{1257343847261}{320320} $ 
& $ -\frac{63166996465891}{2050048} $
 \\[1mm]
$I^7_3$& & & 
& $ -\frac{8}{9} $ 
& $ -\frac{99856}{693} $ 
& $ -\frac{9186824}{1155} $ 
& $ -\frac{1184633742}{5005} $ 
& $ -\frac{68301420326}{15015} $ 
& $ -\frac{491510613571}{8008} $
 \\[1mm]
$I^8_3$& & &  
& $ -\frac{1}{45} $ 
& $ -\frac{47132}{3465} $ 
& $ -\frac{1821254}{1155} $ 
& $ -\frac{832926333}{10010} $ 
& $ -\frac{79252970161}{30030} $ 
& $ -\frac{9152071916597}{160160}$
 \\[1mm]
$I^9_3$& & & &
& $ -\frac{20}{33} $ 
& $ -\frac{592882}{3465} $ 
& $ -\frac{497127053}{30030} $ 
& $ -\frac{25896468179}{30030} $ 
& $ -\frac{13993921905973}{480480}$
 \\[1mm]
$I^{10}_3$& & & &
& $ -\frac{1}{99} $ 
& $ -\frac{351733}{34650} $ 
& $ -\frac{3509496413}{1801800} $ 
& $ -\frac{306154401563}{1801800} $ 
& $ -\frac{17103011002907}{1921920}$
 \\[1mm]
$I^{11}_3$& & & & &
& $ -\frac{758}{2475} $ 
& $ -\frac{4407532}{32175} $ 
& $ -\frac{674753182}{32175} $ 
& $ -\frac{51744893597}{30030}$
 \\[1mm]
$I^{12}_3$& & & & &
& $ -\frac{379}{103950} $ 
& $ -\frac{1263341}{225225} $ 
& $ -\frac{67387787}{40950} $ 
& $ -\frac{78762396103}{360360}$
 \\[1mm]
$I^{13}_3$& & & & & &
& $ -\frac{11816}{96525} $ 
& $ -\frac{55029404}{675675} $ 
& $ -\frac{4156952687}{225225}$
 \\[1mm]
$I^{14}_3$& & & & & &
& $ -\frac{211}{193050} $ 
& $ -\frac{46266791}{18918900} $ 
& $ -\frac{26185035887}{25225200}$
\\[1mm]
$I^{15}_3$& & & & & & &
& $ -\frac{21248}{525525} $ 
& $ -\frac{6689824}{175175}$
 \\[1mm]
$I^{16}_3$& & & & & & &
& $ -\frac{1328}{4729725} $ 
& $ -\frac{1383884}{1576575}$
 \\[1mm]
$I^{17}_3$& & & & & & & &
& $ -\frac{92}{8085}$ 
 \\[1mm]
$I^{18}_3$& & & & & & & &
& $ -\frac{23}{363825}$
\\
\hline
\hline
\label{tab:oct}
\end{tabular}
\end{table*}

\end{appendix}

\small
\bibliographystyle{mn2e}
\bibliography{Lit}

\end{document}

%% file: Befehle.tex
\newcommand{\nbb}{{n^{\rm pl}}}

\newcommand{\vbetac}{{{\boldsymbol\beta}_{\rm c}}}
\newcommand{\vbetach}{{\hat{{\boldsymbol\beta}}_{\rm c}}}
\newcommand{\betac}{\beta_{\rm c}}

\newcommand{\muc}{\mu_{\rm c}}

\newcommand{\vgh}{{\hat{\boldsymbol\gamma}}}
\newcommand{\vghp}{{\hat{\boldsymbol\gamma}'}}

\newcommand{\vbh}{{\boldsymbol{\hat{\beta}}}}

\newcommand{\xc}{x_{\rm c}}

\newcommand{\id}{{\,\rm d}}

\newcommand{\beq}{\begin{equation}}   %

\newcommand{\eeq}{\end{equation}}   %

\newcommand{\beqa}{\begin{eqnarray}}   %

\newcommand{\eeqa}{\end{eqnarray}}   %

\newcommand{\beal}{\begin{align}}

\newcommand{\enal}{\end{align}}

\newcommand{\bspl}{\begin{split}}

\newcommand{\espl}{\end{split}}

\newcommand{\bsub}{\begin{subequations}}

\newcommand{\esub}{\end{subequations}}

\newcommand{\bmulti}{\begin{multline}}   %

\newcommand{\beqm}{\begin{mathletters}}   %

\newcommand{\eeqm}{\end{mathletters}}   %

\newcommand{\me}{m_{\rm e}}

\newcommand{\Ne}{N_{\rm e}}

\newcommand{\Te}{T_{\rm e}}

\newcommand{\Tg}{T_{\gamma}}

\newcommand{\The}{\theta_{\rm e}}

\newcommand{\sigT}{\sigma_{\rm T}}

\newcommand{\vek} [1]{\mbox{\boldmath${#1}$\unboldmath}}

\newcommand{\pot}[2]{#1 \times 10^{#2}}

